\documentclass[letterpaper, journal]{IEEEtran}
\usepackage{algorithm}
\usepackage{multirow}
\usepackage{amsmath,amssymb,amsfonts}
\usepackage{amsthm}
\usepackage{graphicx}
\usepackage{setspace}
\usepackage{multirow}
\usepackage{cite}
\usepackage{float}
\usepackage{color}
\usepackage{alltt, listings}
\usepackage{algpseudocode}
\usepackage{wrapfig}
\usepackage{url}
\usepackage{booktabs} 
\usepackage{bbding}
\usepackage{bm}
\usepackage{enumitem}
\usepackage{hyperref}

\usepackage{xspace}
\newcommand{\model}{\text{Llama4Rec}\xspace}
\newcommand{\lsc}[1]{\textcolor{black}{ #1 }}

\usepackage{multirow}
\usepackage{enumitem}
\usepackage{bm}
\usepackage{subfig}
\usepackage{adjustbox}

\begin{document}

\title{Integrating Large Language Models into Recommendation via Mutual Augmentation and Adaptive Aggregation}


%


\author{Sichun Luo\textsuperscript{*}, Yuxuan Yao\textsuperscript{*}, Bowei He\textsuperscript{*}, Wei Shao\textsuperscript{*}, Jian Xu\textsuperscript{*}, Yinya Huang, Aojun Zhou, Xinyi Zhang\textsuperscript{$\dagger$}, Yuanzhang Xiao, Hanxu Hou\textsuperscript{$\dagger$}, Mingjie Zhan, Linqi Song\textsuperscript{$\dagger$}
\thanks{\textsuperscript{*}Equal Contribution. \textsuperscript{$\dagger$}Corresponding Author.}%
\thanks{Sichun Luo is with the Dongguan University of Technology, City University of Hong Kong, and City University of Hong Kong Shenzhen Research Institute. email: sichunluo2@gmail.com.}%
\thanks{Yuxuan Yao, Wei Shao, Yinya Huang, and Linqi Song are with the City University of Hong Kong and City University of Hong Kong Shenzhen Research Institute.
Bowei He is with the City University of Hong Kong.
}%
\thanks{Jian Xu is with Tsinghua University.}%
\thanks{Aojun Zhou is with the Chinese University of Hong Kong.}%
\thanks{Yuanzhang Xiao is with the University of Hawaii.}%
\thanks{Xinyi Zhang is with the Capital University of Economics and Business.}%
\thanks{Hanxu Hou is with the Dongguan University of Technology.}%
\thanks{Mingjie Zhan is with SenseTime Research.}%
}

\maketitle

\begin{abstract}
Conventional recommender systems and Large Language Model (LLM)-based recommender systems each have their strengths and weaknesses. While conventional recommendation methods excel at mining collaborative information and modeling sequential behavior, they struggle with data sparsity and the long-tail problem. LLM, on the other hand, is proficient at utilizing rich textual contexts but faces challenges in mining collaborative or sequential information. Despite their individual successes, there is a significant gap in leveraging their ensemble potential to enhance recommendation performance.

In this paper, we introduce a general and model-agnostic framework known as \underline{L}arge \underline{la}nguage models with \underline{m}utual augmentation and \underline{a}daptive aggregation for \underline{Rec}ommendation (\textbf{\model}), aiming to bridge this gap via \textit{explicitly ensemble LLM and conventional recommendation model} for more effective recommendation. We propose data augmentation and prompt augmentation strategies tailored to enhance the conventional recommendation model and LLM respectively. An adaptive aggregation module is adopted to combine the predictions of both kinds of models to refine the final recommendation results. Empirical studies on three datasets validate the superiority of \model, demonstrating significant improvements in recommendation performance.
\end{abstract}


\ifCLASSOPTIONcompsoc
\IEEEraisesectionheading{\section{Introduction}}
\else


\section{Introduction}
Recommender systems have emerged as crucial solutions for mitigating the challenge of information overload \cite{zhang2019deep,luo2022personalized,luo2024privacy,luo2024perfedrec++,luo2025rallrec+,luo2025reasoning}. 
Recommender systems encompass a multitude of tasks, such as rating prediction \cite{steck2013evaluation,khan2021deep,luo2024large} and top-$k$ recommendation \cite{le2021efficient,yang2012top,luo2023improving}.
The top-$k$ recommendation, which encompasses collaborative filtering-based direct recommendation \cite{he2017neural,he2020lightgcn}, sequential recommendation \cite{kang2018self,sun2019bert4rec,chen2023sim2rec}, and more, has found wide applications in various areas.
However, recommender systems still suffer from the
\textit{data sparsity} and \textit{long-tail problem}.
Data sparsity arises from sparse user-item interactions, making the task of accurately capturing user preferences more challenging.
The long-tail problem further intensifies the data sparsity issue, as a substantial number of less popular items (\textit{i.e.},  long-tail items) are infrequently interacted with, leading to inadequate data for effective model training and compromised recommendation quality.


\begin{figure}[t!]
    \centering
    \includegraphics[width=1\linewidth]{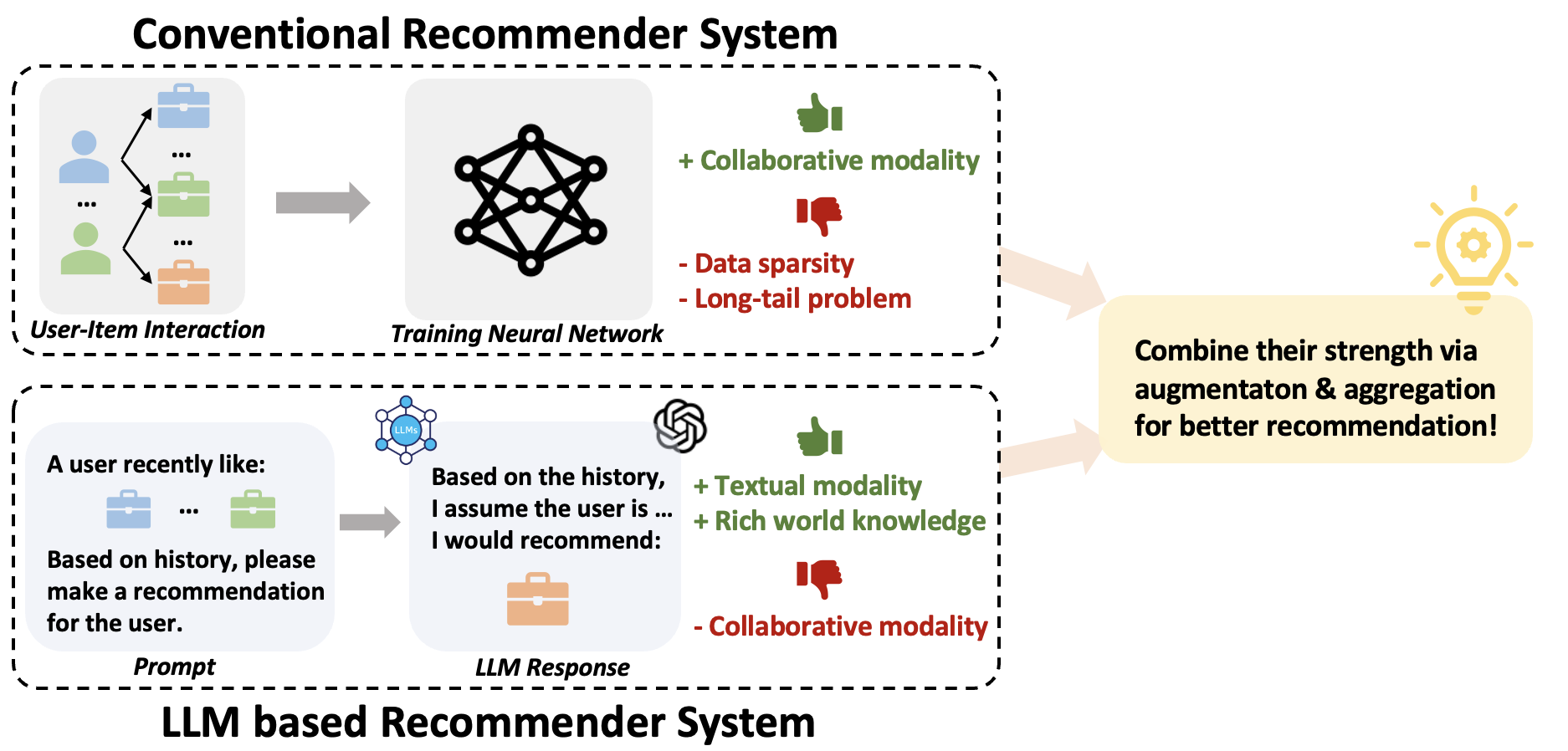}
    \caption{An example that illustrates the motivation of \model.}
    \label{fig:int}
    \vspace{-0.2in}
\end{figure}

In recent years, Large Language Models (LLMs) have emerged, exhibiting exceptional capabilities in language understanding, text generation, and complex reasoning tasks \cite{openai2023gpt4,touvron2023llama,touvron2023llama2,anil2023palm}. Recent studies have started exploring their applicability in recommender systems \cite{liu2023chatgpt,bao2023tallrec,zhang2023recommendation}.
For example,
Liu et al. employed ChatGPT with in-context learning (ICL) for various recommendation tasks \cite{liu2023chatgpt}.
Further progress has been achieved by adopting the instruction tuning technique \cite{ouyang2022training,longpre2023flan} to align general-purpose LLMs with recommendation tasks for improved performance \cite{zhang2023recommendation,bao2023tallrec}.
For instance, TALLRec \cite{bao2023tallrec} reformulates the recommendation problem as a binary classification task and introduces an effective instruction fine-tuning framework for adapting the LLaMA model \cite{touvron2023llama2}. 
However, these LLM-based recommendation methods may not perform optimally as they do not harness the collaborative or sequential information captured by conventional recommendation models. 


Conventional recommendation models and LLM-based recommendation methods each have their respective strengths and weaknesses. 
An illustrative example is shown in Figure \ref{fig:int}.
Conventional methods excel in mining collaborative information and modeling sequential behaviors, while LLMs are proficient in leveraging rich textual contexts.

{Model ensemble} has emerged as a powerful technique in the field of {reliable machine learning} \cite{doppala2022reliable,zounemat2021ensemble}. One of the key advantages of model ensemble is its ability to improve model performance. Each model in the ensemble may capture different aspects of the underlying data patterns, leading to a more comprehensive and robust problem understanding. By aggregating the predictions of these models, we can obtain a final prediction that is often more accurate and reliable than that of any single model.
Inspired by the concept of {model ensemble}, the integration of LLMs into recommender systems presents a significant opportunity to amalgamate the advantages of both methods while circumventing their respective shortcomings. There have been initial attempts to harness the strengths of both conventional and LLM-based recommender systems \cite{geng2022recommendation,zhang2023collm,zheng2023adapting,wei2023llmrec}. Some efforts have sought to integrate collaborative or sequential information by enabling LLMs to comprehend user/item ID information \cite{geng2022recommendation,zhang2023collm,zheng2023adapting}. For instance, a concurrent study by Zhang et al. encodes the semantic embedding into the prompt and sends it to LLM \cite{zhang2023collm}.
\lsc{
Besides, Zheng et al. \cite{zheng2023adapting} proposes adapting large language models for recommendation by integrating collaborative semantics, combining contextual understanding with collaborative filtering to enhance recommendation performance.}
On the other hand, some research works have aimed to augment conventional models using LLMs via data or knowledge augmentation \cite{wei2023llmrec,xi2023towards}. For example, LLMRec \cite{wei2023llmrec} enhances recommender systems by deploying LLMs to augment the interaction graph, thereby addressing the challenges of data sparsity and low-quality side information.







However, existing methods have several limitations.
\textit{Firstly}, current methods lack \textit{generalizability}. Current strategy of integrating ID information in LLMs proves challenging to generalize across different domains and necessitates additional training. Moreover, the current data augmentation method is not universally applicable, as it only addresses a limited number of recommendation scenarios.
\textit{Secondly}, current research primarily focuses on the integration at the data-level (\textit{e.g.}, data augmentation) or model-level (\textit{e.g.}, making LLM understand ID semantics), leaving the result-level integration largely unexplored.
\textit{Lastly}, there is an absence of a comprehensive framework that combines and integrates these methods into a holistic solution.
In light of these limitations, our objective is to explore the integration of conventional recommendation models and LLM-based recommendation methods in depth to address the above limitations and enhance recommendation performance.

In this paper, we introduce a generic framework named \textbf{L}arge \textbf{la}nguage {m}odel with \textbf{m}utual {a}ugmentation and \textbf{a}daptive {a}ggregation for \textbf{Rec}ommendation, referred to as \textbf{\model}, for brevity.
The core idea of \model is to allow conventional recommendation models and LLM-based recommendation models to mutually augment each other, followed by an adaptive aggregation of the augmented models to merge the textual and collaborative modality and yield more optimized results.
Specifically, \model performs data augmentation for conventional recommendation models by leveraging instruction-tuned LLM to alleviate the data sparsity and long-tail problem and thus enrich the collaborative modality. The data augmentation is tailored with different strategies depending on the recommendation scenarios.
Furthermore, we use conventional recommendation models to perform prompt augmentation for LLMs, enhancing the textual modality. The prompt augmentation includes enriching collaborative information from similar users and providing prior knowledge from the conventional recommendation model within the prompt.
We also propose an adaptive aggregation module that merges the predictions of the LLM and conventional models in an adaptive manner. This module is designed as a simple yet effective way to combine the strengths of both models and refine the final recommendation results.
We conduct empirical studies on three real-world datasets, encompassing three different recommendation tasks, to validate the superiority of our proposed method. 
\lsc{The results consistently showcase its superior performance compared to baseline methods, achieving on average 14.65\% improvement in direct recommendation, a 14.21\% improvement in sequential recommendation, and a 3.10\% improvement in rating prediction.}

In a nutshell, the contributions of this work are threefold.
\begin{itemize}[leftmargin=*]
\item 
We introduce \model, a general and model-agnostic framework to integrate LLM into conventional recommendation models, merging the textual modality with collaborative modality. 
\model performs the data augmentation for conventional models to alleviate the data sparsity problem and improve model performance.
The prompt augmentation is applied to LLM for leveraging the information captured by the collaborative-modal recommendation models.
\item 
\model~employs an adaptive aggregation approach to 
combine the prediction from the conventional recommendation model and LLM for improved recommendation performance via leveraging and merging the information captured by such kinds of models working on different modalities.

\item To validate the effectiveness of \model, we conduct extensive experiments on three real-world datasets across three diverse recommendation tasks. The empirical results demonstrate that \model~outperforms existing baselines, exhibiting notable improvements across multiple performance metrics.

\end{itemize}

\section{Related Work}

\subsection{Conventional Recommendation Methods}

Conventional recommendation methods serve as the cornerstone for the contemporary landscape of recommender systems \cite{zhang2019deep}. 
Representative recommendation tasks include rating prediction, collaborative filtering-based direct recommendation, and sequential recommendation, where the latter two are usually formulated as top-$k$ recommendation problems.
Specifically, one of the seminal techniques is the use of matrix factorization for rating prediction, popularized by methods such as Singular Value Decomposition (SVD) \cite{koren2009matrix}.
Collaborative filtering (CF) is another commonly used technique for the recommender systems \cite{adomavicius2005toward}. Recent advancements have evolved CF techniques into more complex neural network architectures and graph-based models \cite{he2017neural,wang2019neural} to enhance the model performance.
On the other hand, sequential recommendation models incorporate temporal patterns into the recommendation pipeline. Techniques such as recurrent neural networks have been adapted for this purpose \cite{hidasi2015session}. Recent research focuses on applying attention mechanisms to further refine these models, leading to a noteworthy boost in performance \cite{sun2019bert4rec,kang2018self}.


Although conventional recommendation techniques are well-suited for capturing latent information associated with users and items, they often require a substantial amount of user-item interactions to provide accurate recommendations, which limits their effectiveness in data sparse and long-tail scenarios \cite{park2008long}.



\begin{figure*}[t]
    \centering
\includegraphics[width=0.98\textwidth]{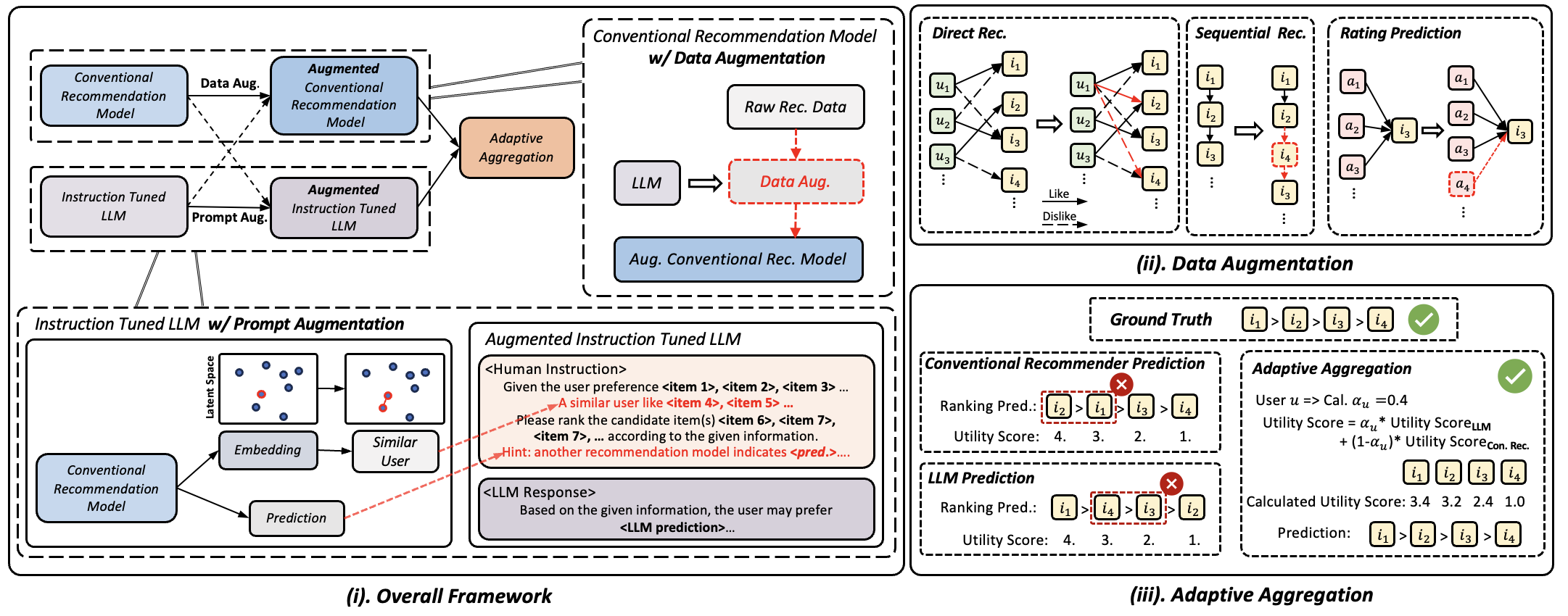}
    \caption{(i) The overall framework architecture of the proposed \model consists of two main components: mutual augmentation and adaptive aggregation. 
    The red dashed lines denoted the data augmentation and prompt augmentation process, respectively.
(ii) Illustration of the data augmentation process, encompassing three diverse recommendation scenarios.
(iii) The pipeline of the adaptive aggregation module, which includes the aggregation of predictions from both types of models.
\lsc{
The symbols $u$, $i$, and $a$ represents user, item, and attributes, respectively.
}
    }
\label{fig:overall}
\vspace{-0.1in}
\end{figure*}

\subsection{Large Language Model for Recommendation}

{\color{black}
Large Language Models (LLMs) have ushered in a paradigm shift across various domains, including recommender systems~\cite{zhao2024recommender,fan2023recommender}. Their ability to perform contextual understanding and reasoning offers significant advantages for recommendation tasks~\cite{lin2025can,dong2022survey}. Existing approaches leveraging LLMs in recommender systems can be broadly categorized into two groups: LLMs as recommenders and LLMs as enhancers.

The first category involves adapting LLMs to directly serve as recommenders, typically through prompting~\cite{dai2023uncovering}, in-context learning~\cite{wang2023zero,liu2023chatgpt}, or fine-tuning~\cite{bao2023tallrec,zhang2023recommendation,luo2024recranker}. For instance, Liu et al.~\cite{liu2023chatgpt} explored the use of ChatGPT across diverse recommendation tasks, demonstrating its effectiveness in specific contexts through rigorous experiments and human evaluations. Similarly, Bao et al.~\cite{bao2023tallrec} proposed TALLRec, which fine-tunes LLMs using instruction tuning datasets tailored to recommendation tasks, aligning LLMs with recommendation objectives.
The second category employs LLMs as enhancers to augment traditional recommender systems by generating or extracting semantic features, such as item descriptions or user profiles~\cite{wei2024llmrec,xi2024towards,wang2024large,zheng2024adapting,ren2024representation}. For example, Wei et al.~\cite{wei2024llmrec} introduced LLMRec, which integrates LLMs with graph augmentation to enrich item representations for recommendation. By leveraging the semantic capabilities of LLMs, these approaches enhance the performance of conventional recommenders, offering a complementary strategy to improve recommendation quality.

In contrast to prior work, which typically adopts either the LLM as recommender or LLM as enhancer approach, our method uniquely combines the strengths of both paradigms. By integrating LLMs as both direct recommenders and enhancers within a unified framework, our approach leverages contextual understanding and semantic feature extraction to achieve superior recommendation performance. 
}

\begin{table}
\centering
\lsc{
\caption{\lsc{Summary of notations used in this work.}}  
\resizebox{0.4\textwidth}{!}{%
\begin{tabular}{l|l}  
\toprule 
\textbf{Notation} & \textbf{Definition} \\
\midrule
$u, i$ & Users and items \\
$\hat{y}_{ui}$ & Predicted score of user $u$ for item $i$ \\
$e_u, e_i$ & Embeddings of user $u$ and item $i$ \\
$\mathcal{D}$ & Training dataset \\
$\mathcal{A}$ & Set of attributes \\
$\mathcal{P}$ & Prompt for the LLM \\
$\mathcal{L}_{\text{BPR}}$ & Bayesian personalized ranking (BPR) loss \\
$\ell_u$ & Long-tail coefficient for user $u$ \\
$N(u)$ & Number of interactions for user $u$ \\
$U$ & Utility score \\
$\Theta$ & Original parameters of the LLM \\
$\alpha_1, \alpha_2$ & Hyperparameters \\
\bottomrule 
\end{tabular}%
}
\label{tab:notation}
}
\end{table}

\section{Methodology}
\subsection{Overview}


Figure~\ref{fig:overall} depicts the architecture of \model, which consists of three components: data augmentation, prompt augmentation, and adaptive aggregation. More specifically,
\model leverages an instruction-tuned LLM to enhance conventional recommender systems through data augmentation. The specific data augmentation strategies for different recommendation situations are detailed in Section~\ref{sec:da}.
In addition, we employ conventional recommendation models to augment the LLM via prompt augmentation, with details in Section~\ref{sec:pa}. 
To further refine the predictions of the conventional model and the LLM, we propose a simple yet effective adaptive aggregation module in Section~\ref{sec:aa}. Lastly, we describe the training strategy for LLM in Section \ref{sec:trainllm}.
\lsc{Summary of notations are listed in Table~\ref{tab:notation}.}






\subsection{Data Augmentation for Conventional Recommender System}
\label{sec:da}

We design ad-hoc data augmentation strategies for different recommendation scenarios to mitigate prevalent issues of data sparsity and the long-tail problem.
This design is motivated by the fact that data distribution and tasks significantly vary across different recommendation scenarios.
In the context of direct recommendation, we capitalize on the power of the instruction-tuned LLM to predict items that a user may like or dislike. We form pairs of these items to calculate the Bayesian Personalized Ranking (BPR) \cite{rendle2012bpr} loss.
For sequential recommendation, we harness the capabilities of the instruction-tuned LLM to predict items that are highly preferred by the user. These predicted items are then randomly inserted into the sequence of items the user has interacted with.
For rating prediction, we utilize the LLM to extract valuable side information (\textit{i.e.}, missing attributes), which is then seamlessly integrated as additional features within the training data.

\subsubsection{\textbf{Data Augmentation for Direct Recommendation}}

For direct recommendation, the Bayesian Personalized Ranking (BPR) loss \cite{rendle2012bpr} is commonly used to optimize the model. The objective of BPR is to maximize the score difference between correctly recommended items and incorrectly recommended items, thereby improving the accuracy of recommendations. The BPR loss is defined~as:
\begin{equation}
\mathcal{L}_{BPR} = -\sum_{(u,i,j) \in \mathcal{D}} \log \sigma(\hat{y}_{ui} - \hat{y}_{uj}), 
\end{equation}
where $(u,i,j)$ refers to a triple of user-item pairs, and the user $u$ has interacted with item $i$ (positive item) and item $j$ (negative item). $\mathcal{D}$ represents the set of such user-item pairs in the training data. $\hat{y}_{ui}$ denotes the predicted score or preference of user $u$ for item $i$.
\lsc{
$\sigma$ is the sigmoid function and 
$\sigma(\hat{y}{ui} - \hat{y}{uj})$
}

Inspired by this, we propose a data augmentation strategy 
that harnesses the capabilities of the
LLM to enhance the direct
recommendation.
Initially, we randomly select pairs of items for a user $ u$ and prompt the LLM to rank each pair based on the user's likely preference. The ranking prediction based on LLM is then combined with the original data and used to train a direct recommendation model. 
Formally, let $(i_{u1}, i_{u2})$ denote a pair of un-interacted items for a user $u$. The LLM is prompted to rank these items, denoted as $i^+, i^- = \textit{LLM}(\mathcal{P}_1)$, where $\mathcal{P}_1$ is the corresponding prompt and $i^+$ is the item preferred over $i^-$. 
The training data $\mathcal{D}$ is updated as $\mathcal{D}' = \mathcal{D} \cup (u,i^+,i^-)$.
The BPR loss is then updated as:
\begin{equation}
\mathcal{L}_{BPR}' = -\sum_{(u,i,j) \in \mathcal{D}'} \log \sigma(\hat{y}_{ui} - \hat{y}_{uj}). 
\end{equation}



\subsubsection{\textbf{Data Augmentation for Sequential Recommendation}}
For sequential recommendation, the data augmentation strategy involves enriching the sequence of interacted items with additional items predicted by the LLM. 
Let's consider a user $u$ with a corresponding sequence of interacted items $\{i_1, ..., i_l\}$. We randomly sample a list of un-interacted items $\{i_{u1}, ..., i_{ut}\}$, and adopt the prompt $\mathcal{P}_2$ to ask the LLM to predict the item most likely to be preferred by the user, denoted as $i_p = \textit{LLM}(\mathcal{P}_2)$. 
This predicted item $i_p$ is then randomly inserted into the user's sequence, resulting in an augmented sequence $\{i_1, ...,i_p, ..., i_l\}$. This sequence is  used to train a more powerful sequential recommendation model.

By including additional items predicted by the LLM, we can enrich the sequence of items for each user, providing a more comprehensive representation of the user's preferences. This, in turn, can enhance the performance of the conventional recommendation model, leading to more accurate recommendations.

\subsubsection{\textbf{Data Augmentation for Rating Prediction}}
In rating prediction tasks, we introduce the use of in-context learning (ICL) in LLMs to provide side information. This is primarily due to the fact that recommendation datasets may contain incomplete information. 
For instance, the popular Movielens dataset \cite{harper2015movielens} lacks information about the \textit{director} of the movies, which can hinder the performance of a conventional rating prediction model.
To mitigate this issue, we leverage the extensive world knowledge contained in an LLM. We prompt the LLM to provide side information, acting as additional attributes for users/items.

Formally, we denote the rating prediction model as $\mathcal{M}_{r}$, and the attribute set as $\mathcal{A} = \{a_1, a_2, ..., a_n\}$, where $a_i \in \mathcal{A}$ denotes a distinct attribute. The model predicts the rating as $\mathtt{Pred} = \mathcal{M}_{r}(\mathcal{A})$.
We then prompt the LLM to provide additional attributes, where the prompt $\mathcal{P}_3$ contains some corresponding examples followed by detailed instructions.
The process is denoted as $\{a_{d1}, a_{d2}, ...\} = \textit{LLM}(\mathcal{P}_3)$. The augmented attribute set is then formed as $\mathcal{A}' = \mathcal{A} \cup \{a_{d1}, a_{d2},...\}$.
The model then predicts the rating using the augmented attribute set, denoted as $\mathtt{Pred}' = \mathcal{M}_{r}(\mathcal{A}')$. This approach allows us to leverage the LLM's world knowledge to enhance the performance of the rating prediction model.



\subsection{Prompt Augmentation for LLM}
\label{sec:pa}

Previous works \cite{bao2023tallrec,zhang2023recommendation} instruction tuning LLM for recommendation in a standard manner. However, these methods can be sub-optimal due to the challenges of distinguishing users based solely on text-based prompt descriptions. Although some concurrent studies \cite{zhang2023collm,zheng2023adapting} incorporate unique identifiers to differentiate users, these approaches require complex semantic understanding of IDs and additional training, limiting their generalizability.

We introduce two text-based prompt augmentation strategies for LLM-based recommendations, \textit{i.e.}, 
we incorporate additional information within the prompt to enhance the model performance.
First, we propose prompt augmentation with similar users, identifying users with analogous preferences to enrich the prompt, thereby enhancing the LLM's ability to leverage collaborative information and generate personalized recommendations. Second, we propose prompt augmentation with conventional recommendation model prediction, providing prior knowledge to guide the LLM toward recommendations that align with user preferences. 
Collectively, these strategies harness the strengths of both LLMs and conventional recommendation models, ensuring generalizability across a wide range of recommendation scenarios.
The illustration of prompt augmentation is underlined in Figure \ref{fig:paa}.

\subsubsection{\textbf{Prompt Augmentation with Collaborative Information from Similar Users}}




To incorporate collaborative information within the prompt and facilitate LLM reasoning, we introduce a prompt augmentation strategy with similar users.
Initially, we utilize a pre-trained conventional recommendation model to acquire embeddings for each user. These embeddings serve as a representation of users in a latent space, which encapsulates their preferences and behaviors.
Specifically, for a user $u$, in conjunction with a conventional recommendation model $\mathcal{M}_c$, we use $\mathcal{M}_c$ to obtain embeddings for each user, denoted as $\{e_1, ..., e_n\}$.
We then calculate the \textit{similarity} between these embeddings in the latent space. 
Various measurements, such as cosine similarity, Jaccard similarity, and Euclidean distance, could be employed in this context. 
In this paper, we calculate the cosine similarity to measure how closely two vectors align, denoted as:
\begin{equation}
sim(u, v) = \frac{e_u \cdot e_v}{||e_u|| \cdot ||e_v||},
\end{equation}
where $e_u$ and $e_v$ are the embeddings of user $u$ and $v$, respectively, and $u,v \in \mathcal{U}$. The $||\cdot||$ denotes the Euclidean norm and $\cdot$ denotes the dot product. 
We identify the pair of users $(u, v)$ that have the highest similarity, indicating that they are the most similar in terms of their preferences and behaviors.
We then use the items interacted with by the most similar user to enrich the prompt for the target user. This strategy leverages the collaborative information gleaned from similar users to generate more relevant and accurate prompts, thereby enhancing the recommendation performance of the LLM.



\subsubsection{\textbf{Prompt Augmentation with Prior Knowledge from Conventional Recommendation Model Prediction}}

To enable the LLM to leverage information captured by conventional models, we propose a prompt augmentation method that incorporates information from conventional recommendation models. 
More specifically,
the augmented prompt is formed by concatenating the original prompt with the prediction from the conventional recommendation model in natural language form.
It's important to note that the prediction from the conventional model varies depending on the recommendation scenarios and base models. 
Through augmenting prompts with predictions from conventional recommendation models, our method integrates collaborative or sequential information captured by these models, thereby enhancing the LLM's contextual understanding and reasoning capabilities and resulting in better recommendation performance.

Notably, unlike ID-based methods such as \cite{zhang2023collm,zheng2023adapting}, our approach relies entirely on text, enabling easy adaptation to new situations.
Moreover, the prompt augmentation could be used as a plug-and-play component for recommendation with closed source LLM, such as the GPT-4 model \cite{openai2023gpt4}.

\subsection{Adaptive Aggregation}
\label{sec:aa}




We endeavor to merge the textual and collaborative modality of both the LLM and conventional recommendation models at the post-learning stage.
Considering the disparate model structures, we 
aggregate the outputs from both types of models for improving recommendation performance.
However, indiscriminate aggregation of model predictions can lead to sub-optimal results.
Conventional recommendation models, known for their susceptibility to the long tail issue, often struggle when dealing with the tail segment. In contrast, LLMs, by leveraging contextual information, are able to maintain a relatively uniform performance across all segments.
Motivated by these observations, 
we first define the long-tail coefficient and subsequently adaptively aggregate the predictions from both model types.

We first define the long-tail coefficient $\ell_u$ for user $u$ to quantify where the user is located in the tail of the distribution. The long-tail coefficient is defined as follows:
\begin{equation}
\ell_{u} = \text{log}\left(N({u})+1\right),
\end{equation}
where $N({u})$ is the number of interaction for user $u$.
A lower long-tail coefficient value indicates that the user has fewer feedback.


While the overarching architecture remains consistent, the implementation details are different for the two tasks considered, namely rating prediction and top-$k$ recommendation.  

\subsubsection{\textbf{Adaptive Aggregation for Rating Prediction}}
For the rating prediction task, we employ an instruction-tuned LLM to predict user-item \textit{utility scores} directly. This approach incorporates the understanding of complex semantics and context by the LLM, which might be overlooked by conventional models.
Similarly, conventional recommendation methods leverage collaborative information and user/item features for predicting the user rating.
Specifically, the LLM is adopted to predict the user rating as {utility score}, denoted as \(U_{LLM}\).
Besides, the conventional predicts the user rating as \(U_{Rec}\).
Subsequently,  
there are various methods to derive a final utility score based on both kinds of models, such as training a neural network to process the utility scores from LLM and conventional models, yielding a final output via learning a complex reflection. However, for the sake of simplicity in this paper, we adopt a simple yet effective linear interpolation approach.
Thus, the final utility score for a user $u$ amalgamates the values from both models, represented as:
\begin{equation}
    U_u =  \alpha_u {U}_{LLM} + (1-\alpha_u) U_{Rec} ,
    \label{eq:1}
\end{equation}
where $\alpha_u$ is the adaptive parameter to control the weight for each model's utility value for user $u$.
We define the $\alpha_u$ as:
\begin{equation}
\alpha_{u} = \text{max} \left( \frac{\ell_{max} - \ell_{u}}{\ell_{max}-\ell_{min}}, \alpha_2 \right ) \cdot \alpha_1,
\label{equ:1}
\end{equation}
where $\ell_{max}$ and $\ell_{min}$ are the maximum and minimum long-tail coefficients of the users, respectively, $\alpha_1$ is a hyper-parameter that controls the weight, and $\alpha_{2} < 1$ is a cut-off weight. 
From Equation \eqref{equ:1}, we can observe that for user \( u \), the further they are positioned in the long tail (\textit{i.e.}, the fewer items they have interacted with), the lower is the value of \( \ell_u \) and the higher is the value of \( \alpha_u \). As a result, in Equation \eqref{eq:1}, the weight of the utility score from the LLM model becomes more pronounced. This aligns with the motivation we previously discussed.


\subsubsection{\textbf{Adaptive Aggregation for Top-$k$ Recommendation}}

For the top-$k$ recommendation task, the LLM is employed to re-rank the item list generated by a conventional recommendation model.
Specifically, from conventional recommendation methods, we curate a top-ranked list comprising \(k'\) items, denoted as \(\{i_1, ..., i_{k'}\}\). Each item in this list is assigned a utility weight, \({U}^i_{Rec} = -s \cdot \mathcal{C}\), where \(\mathcal{C}\) is a constant and \(s\) represents the position of item \(i\), \textit{i.e.}, $s \in \{1, ... , k'\}$.
A higher utility weight indicates a stronger inclination of the user's preference.
For listwise comparison conducted by the LLM,  the process begins by using the LLM to directly output the predicted order of these candidate items.
Then we assign utility scores for items at each position, denoted as $U_1, {U}_2, ..., U_{k'}$, where $U_1 \ge U_2 \ge ... \ge U_{k'}$.
In this paper, we also set \({U}_{p} = -p \cdot \mathcal{C}\) where ${p} \in \{1, 2, ..., {k'}\}$,
\lsc{and ${U}{p}$ is the utility score assigned to the item at position $p$.}
The final utility score for an item amalgamates the values from both the original rating and the LLM prediction, similar to Equation~\eqref{eq:1}.










\subsection{Training Strategy for LLM}
\label{sec:trainllm}

\subsubsection{\textbf{Instruction Tuning Dataset Construction}}

This section details the creation of an instruction-tuning dataset that encompasses two types of recommendation tasks catering to top-$k$ recommendation and rating prediction scenarios. A depiction of these two tasks, specifically referred to as \textit{listwise ranking} and \textit{rating prediction}, can be found in the Figure \ref{fig:paa}. It is noteworthy that we also employ the LLM to execute \textit{pointwise ranking} within top-$k$ recommendation scenarios, \textit{i.e.}, utilizing LLM to predict ratings for each item within the top-\textit{k} recommendations and sorting the predicted ratings to derive the final result.
\lsc{In our dataset, the input $x$ consists of user and item features (\textit{e.g.}, user preferences, item attributes) used to prompt the model, while the output $y$ represents the corresponding recommendation or prediction. The ground-truth information from the training set is used to construct the output $y$, ensuring that the model learns from accurate and verified data.}

\begin{figure}[h]
    \centering
    \includegraphics[width=1\linewidth]{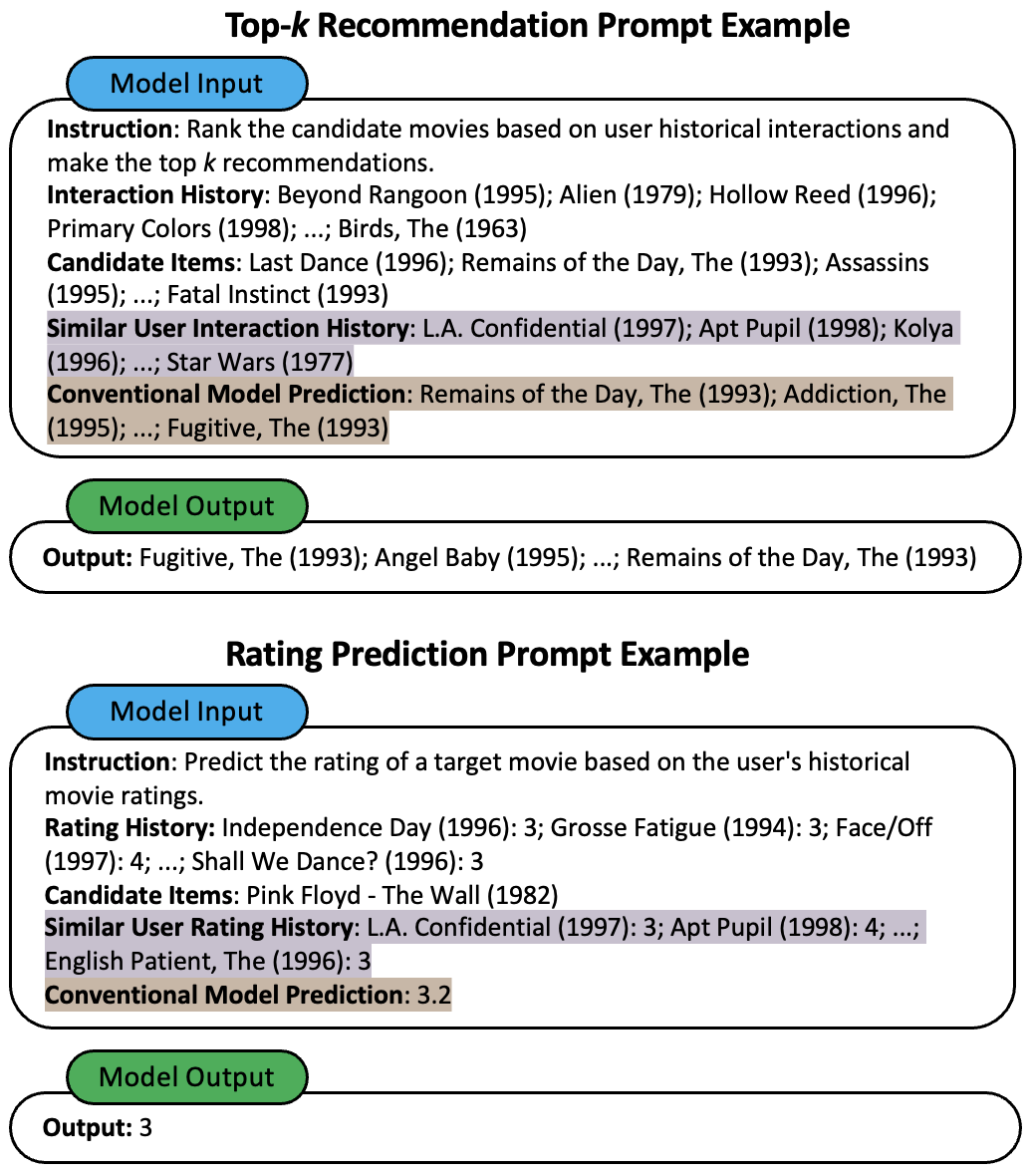}
    \caption{Examples of instructions for top-\textit{k} recommendation and rating prediction. 
The {prompt augmentation} component is highlighted.
\lsc{The similar user histories and conventional model predictions are prompt augmented, which are highlighted in light purple and beige.
We keep the prompt template structure and convert the bullet-point format into cohesive sentences for use in experiments. In the top-k recommendation, both direct and sequential recommendations follow the same prompt structure, with the key difference being that the user interaction history for sequential recommendations is time-series aware.
}
}
    \label{fig:paa}
\end{figure}

\subsubsection{\textbf{Optimization via Instruction Tuning}}
In this work, we perform full parameter instruction tuning to optimize LLMs using generated instruction data.
Due to our need for customization, we chose LLaMA-2 \cite{touvron2023llama2}, an open-source, high-performing LLM, which permits task-specific fine-tuning.
During supervised fine-tuning, we apply a standard cross-entropy loss following Alpaca~\cite{alpaca}.
The training set \( \mathcal{D}_{ins} \) consists of instruction input-output pairs \((x, y)\), which have been represented in natural language. The objective is to fine-tune the pre-trained LLM by minimizing the cross-entropy loss, formalized as:
\begin{equation}
\min _{\Theta} \sum_{(x, y) \in \mathcal{D}_{ins}} \sum_{t=1}^{|y|} -\log P_{\Theta}\left(y_{t} \mid x, y_{[1:t-1]}\right),
\end{equation}
where \( \Theta \) are the original parameters for LLM, $P_{\Theta}$ is the conditional probability, $|y|$ is the number of tokens in $y$, \( y_t \) is the \( t \)-th token in the target output \( y \), and \( y_{[1:t-1]} \) represents tokens preceding \( y_t \) in \( y \). 
By minimizing this loss function, the model fine-tunes its parameters \( \Theta \) to adapt to the specifics of the new instruction tuning dataset \( \mathcal{D}_{ins} \), while leveraging the general language understanding and reasoning that has been acquired during pre-training \cite{zhang2023instruction}. In this manner,
LLM can capture the user's preferences for items expressed in natural language, facilitating diverse recommendation tasks, including top-\textit{k} recommendation and rating prediction.

\begin{table*}[h]
\centering
\caption{ Performance achieved by different direct recommendation methods. \lsc{H@$k$ indicates the hit rate at position $k$, and N@$k$ represents normalized discounted cumulative gain (NDCG) at position $k$.}
 }
\label{tab:t2}
\begin{adjustbox}{width=\linewidth}
\begin{tabular}{ccc c c ccccccccccc}
\toprule

\multirow{2}{*}{Backbone} & \multirow{2}{*}{Method} & \multicolumn{4}{c}{ML-100K}  & \multicolumn{4}{c}{ML-1M}& \multicolumn{4}{c}{BookCrossing}  \\ \cmidrule(lr){3-6}\cmidrule(lr){7-10}\cmidrule(lr){11-14}
\multirow{2}{*}{}&\multirow{2}{*}{}& H@3 $\uparrow$ &  N@3 $\uparrow$& H@5 $\uparrow$ &  N@5 $\uparrow$& H@3 $\uparrow$ &  N@3 $\uparrow$& H@5 $\uparrow$ &  N@5 $\uparrow$& H@3 $\uparrow$ &  N@3 $\uparrow$& H@5 $\uparrow$ &  N@5 $\uparrow$ \\ \midrule

\multirow{4}{*}{MF} & Base &0.0455 & 0.0325 &0.0690 & 0.0420&0.0255 & 0.0187 &0.0403 &0.0248 & 0.0294 & 0.0227 &0.0394 & 0.0269\\ 
\multirow{4}{*}{}& IFT &0.0546 & 0.0388 &0.0790 & 0.0488 &0.0242 & 0.0175 &0.0410 & 0.0244&0.0247 & 0.0177 &0.0377 & 0.0230\\
\multirow{4}{*}{}& \model &\textbf{0.0645*}&\textbf{0.0474*}&\textbf{0.0919*}&\textbf{0.0588*}&\textbf{0.0281*}&\textbf{0.0203*}&\textbf{0.0433*}&\textbf{0.0265*}&\textbf{0.0365*}&\textbf{0.0284*}&\textbf{0.0462*}&\textbf{0.0324*}\\
\multirow{4}{*}{}& Impro.  & 18.13\%&22.16\%&16.33\%&20.49\%&10.20\%&8.56\%&5.61\%&6.85\%&24.15\%&25.55\%&17.26\%&20.45\%
\\
\midrule
\multirow{4}{*}{LightGCN} & Base & {0.0492} & 0.0343 &0.0744 & 0.0447 & 0.0283 & 0.0203 &0.0432 & 0.0264&0.0358 & 0.0272 &0.0480 & 0.0322\\ 
\multirow{4}{*}{} & IFT & 0.0537 & 0.0381 &0.0846 & 0.0507 &0.0268 & 0.0193 &0.0441 & 0.0263&0.0287 & 0.0202 &0.0448 & 0.0268\\
\multirow{4}{*}{} & \model &\textbf{0.0647*}&\textbf{0.0476*}&\textbf{0.0967*}&\textbf{0.0608*}&\textbf{0.0304*}&\textbf{0.0222*}&\textbf{0.0461*}&\textbf{0.0286*}&\textbf{0.0434*}&\textbf{0.0338*}&\textbf{0.057*}&\textbf{0.0394*}\\
\multirow{4}{*}{}& Impro. &  20.48\%&24.93\%&14.30\%&19.92\%&7.42\%&9.36\%&4.54\%&8.33\%&21.23\%&24.26\%&18.75\%&22.36\%
 \\\midrule
\multirow{4}{*}{MixGCF} & Base &0.0526 & 0.0401 &0.0757 & 0.0496 &0.0159 & 0.0115 &0.0238 & 0.0147&0.0426 & 0.0330 &0.0556 & 0.0384\\ 
\multirow{4}{*}{} & IFT &0.0617 & 0.0452 &0.0906 & 0.0570 &0.0162 & 0.0114 &\textbf{0.0259} & 0.0154& 0.0337 & 0.0243 &0.0506 & 0.0312 \\
\multirow{4}{*}{} & \model &\textbf{0.0690*}&\textbf{0.0515*}&\textbf{0.0949*}&\textbf{0.0621*}&\textbf{0.0174*}&\textbf{0.0128*}&\textbf{0.0259}&\textbf{0.0162*}&\textbf{0.0495*}&\textbf{0.0384*}&\textbf{0.0635*}&\textbf{0.0441*}\\
\multirow{4}{*}{}& Impro. &  11.83\%&13.94\%&4.75\%&8.95\%&7.41\%&11.30\%&0.00\%&5.19\%&16.20\%&16.36\%&14.21\%&14.84\%
 \\\midrule
\multirow{4}{*}{SGL} & Base & 0.0505 & 0.0380 &0.0729 & 0.0472 &0.0284 & 0.0206 &0.0434 & 0.0267& 0.0419 & 0.0319 &0.0566 & 0.0380\\ 
\multirow{4}{*}{} & IFT & 0.0520 & 0.0392 &0.0792 & 0.0503 &0.0275 & 0.0202 &0.0438 & 0.0269& 0.0326 & 0.0237 &0.0499 & 0.0307\\
\multirow{4}{*}{} & \model & \textbf{0.0632*}&\textbf{0.0479*}&\textbf{0.0917*}&\textbf{0.0596*}&\textbf{0.0308*}&\textbf{0.0224*}&\textbf{0.0480*}&\textbf{0.0294*}&\textbf{0.0501*}&\textbf{0.0393*}&\textbf{0.0634*}&\textbf{0.0448*}\\
\multirow{4}{*}{}& Impro. & 21.54\%&22.19\%&15.78\%&18.49\%&8.45\%&8.74\%&9.59\%&9.29\%&19.57\%&23.20\%&12.01\%&17.89\%
 \\

\bottomrule
\end{tabular}
\end{adjustbox}
\end{table*}

\section{Experiment}

In this section, we present a thorough empirical evaluation to validate the effectiveness of our proposed framework.
Specifically, our objective is to investigate whether the incorporation of our proposed \model~can enhance existing recommendation models.
The overarching goal is to answer the following research questions:


\begin{itemize}[leftmargin=*]
\item
\textbf{RQ1}: Does our proposed \model~framework enhance the performance of existing recommendation models?
\item
\textbf{RQ2}: How do the various modules in \model affect the recommendation performance?
\item
\textbf{RQ3}: How do different hyper-parameters impact the overall performance of the framework?
\textbf{RQ4}: How does the instruction-tuned model compare to other LLMs, such as GPT? How would the model perform \textit{w.r.t.} model and data scaling?
\end{itemize}

\subsection{Experiment Setup}



\subsubsection{\textbf{Dataset}}
Following \cite{bao2023tallrec}, we rigorously evaluate the performance of our proposed framework by employing three heterogeneous, real-world datasets. 
\textbf{MovieLens}\footnote{\url{https://grouplens.org/datasets/movielens/}} \cite{harper2015movielens} serve as benchmark datasets in the realm of movie recommendation. We employ two variants of the dataset: MovieLens-100K (\textbf{ML-100K}) and MovieLens-1M (\textbf{ML-1M}). The former consists of approximately 100,000 user-item ratings, while the latter scales up to roughly 1,000,000 ratings.
\textbf{BookCrossing}\footnote{Due to the absence of timestamp data, we synthesize historical interactions through random sampling.} \cite{ziegler2005improving} includes user-generated book ratings on a scale of 1 to 10, alongside metadata such as ‘Book-Author' and ‘Book-Title'.
We employ the LLM to augment the ‘director' and ‘star' features for ML-100K and ML-1M datasets, and augment the ‘genre' and ‘page length' features for the BookCrossing dataset. To ensure the data quality, we adopt the 5-core setting, where we filter unpopular users and items with fewer than five interactions for the BookCrossing dataset.
The key characteristics of these datasets are delineated in the Table \ref{dataset}.

\begin{table}[t]
\centering
\caption{Dataset Description.}
\label{dataset}
\begin{adjustbox}{width=\linewidth}
\begin{tabular}{c c c c c}
\toprule   
&ML-100K&ML-1M&BookCrossing\\\midrule
\# of User&943&6,040&6,851\\
\# of Item&1,682&3,706&9,085\\
\# of Rating&100,000&1,000,209&115,219\\
Density&0.063046&0.044683&0.001851\\\midrule
\multirow{2}{*}{User Features}& Gender, ZipCode,&Gender, ZipCode, &\multirow{2}{*}{Location, Age}\\
\multirow{2}{*}{} & Occupation, Age&Occupation, Age&\multirow{2}{*}{}\\\midrule
\multirow{2}{*}{Item Features}& {Title, Genres}& \multirow{2}{*}{Title, Genres}&Title, Author,  \\  
\multirow{2}{*}{}&Year&\multirow{2}{*}{}&Year, Publisher\\\midrule
{Augmented}& {Movie Director},& {Movie Director},&{Book Genres},\\
{Features}& {Movie Star} & {Movie Star} &{Page Length}\\

\bottomrule
\end{tabular}
\end{adjustbox}

\end{table}

\subsubsection{\textbf{Evaluation Metrics.}}
Aligning with \cite{he2020lightgcn,sun2019bert4rec}, for the top-$k$ recommendation task, we turn to two well-established metrics: Hit Ratio (HR) and Normalized Discounted Cumulative Gain (NDCG), denoted by H and N, respectively. In our experiments, $k$ is configured to be either 3 or 5 for a comprehensive evaluation, similar to the experiment setting in \cite{zhang2023recommendation}.
\lsc{For instance, H@$k$ refers to the hit rate at position $k$, while N@$k$ denotes the normalized discounted cumulative gain (NDCG) at position $k$.}
In accordance with \cite{fan2019graph}, we employ Root Mean Squared Error (RMSE) and Mean Absolute Error (MAE) as evaluation metrics for the rating prediction task.

\begin{table*}[h]
\centering
\caption{ Performance achieved by different sequential recommendation methods. \lsc{H@$k$ indicates the hit rate at position $k$, and N@$k$ represents normalized discounted cumulative gain (NDCG) at position $k$.}
 }
\label{tab:t3}
\begin{adjustbox}{width=\linewidth}
\begin{tabular}{ccc c c ccccccccccc}
\toprule

\multirow{2}{*}{Backbone} & \multirow{2}{*}{Method} & \multicolumn{4}{c}{ML-100K}  & \multicolumn{4}{c}{ML-1M}& \multicolumn{4}{c}{BookCrossing}  \\ \cmidrule(lr){3-6}\cmidrule(lr){7-10}\cmidrule(lr){11-14}
\multirow{2}{*}{}&\multirow{2}{*}{}& H@3 $\uparrow$ &  N@3 $\uparrow$& H@5 $\uparrow$ &  N@5 $\uparrow$& H@3 $\uparrow$ &  N@3 $\uparrow$& H@5 $\uparrow$ &  N@5 $\uparrow$& H@3 $\uparrow$ &  N@3 $\uparrow$& H@5 $\uparrow$ &  N@5 $\uparrow$ \\ \midrule

\multirow{4}{*}{SASRec} & Base & 0.0187 & 0.0125 &0.0385 & 0.0205 &0.0277 & 0.0165 &0.0502 & 0.0257&0.0086 & 0.0049 &0.0163 & 0.0081 \\ 
\multirow{4}{*}{} & IFT & 0.0204 & 0.0136 &0.0379 & 0.0207 &0.0241 & 0.0159 &0.0473 & 0.0254&0.0124 & 0.0086 &0.0185 & 0.0111\\
\multirow{4}{*}{}& \model &\textbf{0.0238*}&\textbf{0.0155*}&\textbf{0.0449*}&\textbf{0.0240*}&\textbf{0.0293*}&\textbf{0.0201*}&\textbf{0.0504}&\textbf{0.0287*}&\textbf{0.0142*}&\textbf{0.0098*}&\textbf{0.0227*}&\textbf{0.0131*}\\
\multirow{4}{*}{}&Impro.&16.67\%&13.97\%&16.62\%&15.94\%&5.78\%&21.82\%&0.40\%&11.67\%&14.52\%&13.95\%&22.70\%&18.02\%
  \\\midrule
\multirow{4}{*}{BERT4Rec} & Base & 0.0153 & 0.0104 &0.0294 & 0.0161 &0.0107 & 0.0069 &\textbf{0.0211} & 0.0112&0.0088 & 0.0058 &0.0161 & 0.0088\\ 
\multirow{4}{*}{} & IFT & 0.0174 & 0.0119 &0.0326 & 0.0100 &0.0106 & 0.0071 &0.0188 & 0.0104& 0.0127 & 0.0092 &0.0180 & 0.0113 \\
\multirow{4}{*}{} & \model & \textbf{0.0198*}&\textbf{0.0134*}&\textbf{0.0332}&\textbf{0.0189*}&\textbf{0.0115*}&\textbf{0.0078*}&{0.0206}&\textbf{0.0115*}&\textbf{0.0154*}&\textbf{0.0108*}&\textbf{0.023*}&\textbf{0.0139*}\\
\multirow{4}{*}{}& Impro. & 13.79\%&12.61\%&1.84\%&17.39\%&7.48\%&9.86\%&-2.37\%&2.68\%&21.26\%&17.39\%&27.78\%&23.01\%
 \\\midrule
\multirow{4}{*}{CL4SRec} & Base & 0.0243 & 0.0143 &0.0436 & 0.0222 &0.0259 & 0.0153 &\textbf{0.0492} & 0.0248&0.0083 & 0.0048 &0.0165 & 0.0082\\ 
\multirow{4}{*}{} & IFT & 0.0230 & 0.0149 &0.0428 & 0.0230  &0.0234 & 0.0155 &0.0447 & 0.0241& 0.0102 & 0.0071 &0.0177 & 0.0102\\
\multirow{4}{*}{} & \model  &\textbf{0.0255*}&\textbf{0.0182*}&\textbf{0.0440}&\textbf{0.0255*}&\textbf{0.0278*}&\textbf{0.0185*}&{0.0482}&\textbf{0.0268*}&\textbf{0.0138*}&\textbf{0.0093*}&\textbf{0.0220*}&\textbf{0.0127*}\\
\multirow{4}{*}{}& Impro. &  4.94\%&22.15\%&0.92\%&10.87\%&7.34\%&19.35\%&-2.03\%&8.06\%&35.29\%&30.99\%&24.29\%&24.51\%
 \\

\bottomrule
\end{tabular}
\end{adjustbox}
\end{table*}

\subsubsection{\textbf{Data Preprocessing.}}
Following prior works~\cite{zhang2023recommendation,luo2022hysage}, we adopt a leave-one-out evaluation strategy. More specifically, within each user's interaction sequence, we choose the most recent item as the test instance. The item immediately preceding this serves as the validation instance, while all remaining interactions are used to constitute the training set. Moreover, 
regarding the instruction-tuning dataset construction, 
\lsc{
we randomly sampled 5K instructions for each of the three recommendation tasks (rating prediction, direct recommendation, and sequential recommendation) across
three datasets (ML-100K, ML-1M, and BookCrossing), resulting in a total of 3 × 3 × 5K = 45K raw data.} 
We eliminated instructions that were repetitive or of low quality (identified by users with fewer than three interactions in their interaction history), leaving approximately 25K high-quality instructions.
These instructions are mixed to create an instruction-tuning dataset to fine-tune the LLM.

\subsubsection{\textbf{Backbone Models}}

We incorporate our \model~with the following recommendation models that are often used for various recommendation tasks as the backbone models: 
\begin{itemize}[leftmargin=*]
    
    \item \textbf{\textit{Direct Recommendation.}}  In the scenario of direct recommendation,
    We adopt four representative methods, including:
\textbf{MF} \cite{koren2009matrix},
\textbf{LightGCN} \cite{he2020lightgcn},
\textbf{MixGCF} \cite{huang2021mixgcf}, and
\textbf{SGL} \cite{wu2021self}.
\item \textbf{\textit{Sequential Recommendation.}} 
     Regarding sequential recommendation, we opt for three widely used models, including:
\textbf{SASRec} \cite{kang2018self},
\textbf{BERT4Rec} \cite{sun2019bert4rec}, and
\textbf{CL4SRec} \cite{xie2022contrastive}.

\item \textbf{\textit{Rating Prediction.}} We consider the following classical models for rating prediction, including: 
    \textbf{DeepFM} \cite{guo2017deepfm}, \textbf{NFM} \cite{he2017neural}, \textbf{DCN} \cite{wang2017deep}, \textbf{AFM} \cite{xiao2017attentional}, \textbf{xDeepFM} \cite{lian2018xdeepfm}, and \textbf{AutoInt} \cite{song2019autoint}.
\end{itemize}




We employ the LLaMA-2 7B version as the backbone LLM across all experiments, unless specifically mentioned otherwise.
Our primary comparison is with the standard \textbf{I}nstruction \textbf{F}ine-\textbf{T}uning (\textbf{IFT}) method adopted in TALLRec \cite{bao2023tallrec} and InstructRec \cite{zhang2023recommendation}. For the rating prediction task, LLaMA-2 with IFT is used to directly predict the rating. For the top-k recommendation task, the tuned LLM is used to re-rank the list predicted by the backbone model, in accordance with \cite{hou2023large}, referred to as listwise ranking. Besides, we also adopt LLM for predicting the rating for each item and sort by the predicted scores, referred to as pointwise ranking.  

\subsubsection{\textbf{Implementation Details.}}
During training for LLaMA 2 (7B) with full-parameter tuning, we use a uniform learning rate of $2\times 10^{-5}$ and a context length of 2048, and we set the batch size as 16. Additionally, we use a cosine scheduler for three epochs in total with a 50-step warm-up period. To efficiently train the computationally intensive models, we simultaneously employ DeepSpeed training with ZeRO-3 stage~\cite{rajbhandari2020zero} and flash attention~\cite{dao2022flashattention}. We trained the 7B model on 16 NVIDIA A800 80GB GPUs. For the inference stage, we employed the vLLM framework \cite{kwon2023efficient} with greedy decoding, setting the temperature to 0. Only one GPU was utilized during the inference phase.
We only evaluate the instruction-tuned LLaMA model for rating prediction task since it is not applicable for directly making top-$k$ recommendations.


We implement the models for rating prediction task using the DeepCTR-Torch\footnote{\url{https://github.com/shenweichen/DeepCTR-Torch}} library. For the top-$k$ recommendation task, we utilize the SELFRec\footnote{\url{https://github.com/Coder-Yu/SELFRec}} library \cite{yu2023self} for implementation.
As for the hyper-parameter settings, $\alpha_1$ and $\alpha_2$ are selected from $\{$0.1, 0.3, 0.5, 0.7, 0.9$\}$ respectively for all experiments. 
$\mathcal{C}$ is fixed to 1 and $k'$ is set to 10.
We repeat the experiment five times and calculate the average.
We report the best results obtained when the ranking method is selected from pointwise and listwise ranking.
For all experiments, the best results are highlighted in \textbf{boldfaces}. 
* indicates the statistical significance for
$p \le 0.05$ compared to the best baseline method based on the paired t-test. \textit{Improv.}
denotes the improvement of our method over the best baseline method.




\subsection{Main Results (RQ1)}
We conducted an extensive evaluation of our proposed \model and the baseline methods on three datasets to assess the model's performance under diverse recommendation scenarios. 
The experiment results for \lsc{direct recommendation, sequential recommendation, and rating prediction are shown in Table \ref{tab:t2}, Table \ref{tab:t3}, and Table \ref{tab:t1}, respectively.}
We have the following key observations.

\begin{itemize}[leftmargin=*]
\item
\lsc{As shown in Table \ref{tab:t2}, }\model~consistently outperforms baseline methods in almost all scenarios, with particularly significant improvements observed in the direct recommendation task. 
Moreover, our findings reveal that direct instruction fine-tuning LLMs for recommendation tasks does not consistently yield promising performance. 
These results highlight the effectiveness of integrating LLMs into conventional recommendation models, underscoring the importance of incorporating the mechanism that utilizes instruction-tuned LLM to mutually augment and adaptively aggregate with conventional recommendation models.
\lsc{
\item
We observe a performance decline when dealing with a larger number of candidates (i.e., $ k $) in direct and sequential recommendation scenario, as the improvement for $ k=5 $ is usually less than for $ k=3 $. This suggests a limitation of the LLM, where ranking and recommending the correct items become more challenging as the candidate pool increases. This cause the performance drops in some cases.
}
{\color{black}
This may be attributed to the “lost-in-the-middle” problem \cite{liu2024lost}, where LLMs tend to focus more on the beginning and ending of sequences while neglecting the middle portion. This behavior contributes to a decline in performance as the size of the candidate set increases.
Besides, larger candidate sets inherently increase the complexity of the task, requiring the model to distribute its attention across more options. Smaller or less powerful LLMs may struggle to maintain this balance, leading to reduced performance.
To address this issue, we propose several possible remedies. First, employing more powerful LLMs can help mitigate the problems associated with larger candidate sets, as these models are better equipped to handle longer sequences. Second, Splitting large candidate sets into smaller subsets (e.g., partitioning large $k$ into smaller chunks) and aggregating the results can help the model maintain focus and improve accuracy for higher $k$ values.
}
\item
In the scenario of the rating prediction task in Table~\ref{tab:t1}, while the instruction-tuned LLaMA model significantly underperforms when compared to conventional recommendation models, integrating the LLM yields a marked performance improvement. This suggests that the LLM and conventional recommendation models learn distinct aspects of information. 
Consequently, integrating the LLM with conventional recommendation models could enhance recommendation performance.

\item
In the context of \lsc{direct recommendation and sequential recommendation} recommendations \lsc{in Table \ref{tab:t2} and \ref{tab:t3}}, \model exhibits a more pronounced improvement for direct recommendations task. 
In addition, a more significant enhancement is observed on the Bookcrossing dataset, which can be attributed to the more fine-grained and distinguishable rating of the Bookcrossing dataset.
\item
Besides, we discuss the \textbf{computational efficiency}.
Additional training with augmented data is required in the \model framework, which presents an extra cost. In the current experimental setup, we train a new model from scratch. However, this process could be optimized by continuing to train a previously tuned model, thereby reducing time costs.
Additionally, in our experiment, we observed that training the LLaMA-2 7B model with around 25K instructions on 16 A800 GPUs with 2500 steps took approximately 1.94 hours. The inference time for each instruction averaged about 17 instructions per second, translating to a requirement of around 0.059 seconds per item for computation by a single NVIDIA A800~GPU. 


\end{itemize}


\begin{table}[t]
\centering
\caption{ Performance achieved by different methods in rating prediction task. 
}
\label{tab:t1}
\begin{adjustbox}{width=\linewidth}
\begin{tabular}{ccc c c c ccc}
\toprule

\multirow{2}{*}{Backbone} & \multirow{2}{*}{Method} & \multicolumn{2}{c}{ML-100K}  & \multicolumn{2}{c}{ML-1M}& \multicolumn{2}{c}{BookCrossing}  \\ \cmidrule(lr){3-4}\cmidrule(lr){5-6}\cmidrule(lr){7-8}
\multirow{2}{*}{}&\multirow{2}{*}{}& RMSE $\downarrow$ &  MAE $\downarrow$&  RMSE $\downarrow$ & MAE $\downarrow$&  RMSE $\downarrow$ & MAE $\downarrow$  \\ \midrule
{LLaMA} & IFT & 1.2792 &0.8940 & 1.2302& 0.8770 & 2.0152 &1.3782\\ \midrule
\multirow{3}{*}{DeepFM} & Base & 1.0487 & 0.8082 &  0.9455 & 0.7409 & 1.7738 & 1.3554 \\ 
\multirow{3}{*}{}& \model & \textbf{1.0306*}&\textbf{0.7987*}&\textbf{0.9360*}&\textbf{0.7321*}&\textbf{1.6958*}&\textbf{1.2843*} \\
\multirow{3}{*}{}& Impro.&1.73\%&1.18\%&1.00\%&1.19\%&4.40\%&5.25\%  \\
\midrule
\multirow{3}{*}{NFM} & Base & 1.0284 & 0.8005 & 0.9438 & 0.7364 & 2.121 & 1.5984 \\ 
\multirow{3}{*}{} & \model & \textbf{1.0189*}&\textbf{0.7961}&\textbf{0.9369*}&\textbf{0.7303*}&\textbf{1.9253*}&\textbf{1.4473*}
 \\
\multirow{3}{*}{}& Impro. &0.92\%&0.55\%&0.73\%&0.83\%&9.23\%&9.45\% \\
\midrule
\multirow{3}{*}{DCN} & Base & 1.0478 & 0.8063 &  0.9426 & 0.7342 & 2.0216 & 1.4622\\ 
\multirow{3}{*}{} & \model & \textbf{1.0367*}&\textbf{0.8033}&\textbf{0.9345*}&\textbf{0.7272*}&\textbf{1.8518*}&\textbf{1.3566*}
\\
\multirow{3}{*}{}& Impro. & 1.06\%&0.37\%&0.86\%&0.95\%&8.40\%&7.22\% \\
\midrule
\multirow{3}{*}{AFM} & Base & 1.0471 & 0.8035 & 0.9508 & 0.7464 & 1.6516 & 1.2614\\ 
\multirow{3}{*}{} & \model & \textbf{1.0340*}&\textbf{0.7996}&\textbf{0.9426*}&\textbf{0.7394*}&\textbf{1.6244*}&\textbf{1.2259*}
\\
\multirow{3}{*}{}& Impro. & 1.25\%&0.49\%&0.86\%&0.94\%&1.65\%&2.81\%
 \\
\midrule
\multirow{3}{*}{xDeepFM} & Base & 1.1472 & 0.8836 & 0.9519 & 0.7428 & 2.1756 & 1.6461\\ 
\multirow{3}{*}{} & \model & \textbf{1.0947*}&\textbf{0.8483*}&\textbf{0.9401*}&\textbf{0.7336*}&\textbf{1.9610*}&\textbf{1.4833*}\\
\multirow{3}{*}{}& Impro. & 4.58\%&4.00\%&1.24\%&1.24\%&9.86\%&9.89\%
\\
\midrule
\multirow{3}{*}{AutoInt} & Base & 1.0500 & 0.8120 &  0.9471 & 0.7404 & 1.9148 & 1.4501 \\ 
\multirow{3}{*}{} & \model & \textbf{1.0369*}&\textbf{0.8059*}&\textbf{0.9382*}&\textbf{0.7326*}&\textbf{1.7917*}&\textbf{1.3492*}
\\
\multirow{3}{*}{}& Impro. &1.25\%&0.75\%&0.94\%&1.05\%&6.43\%&6.96\% \\
\bottomrule
\end{tabular}
\end{adjustbox}

\end{table}

\begin{table}[t]
\centering
\caption{ 
Ablation study on key components of \model on the ML-1M and Bookcrossing dataset.
}
\label{tab:ablation}
\begin{adjustbox}{width=1.0\linewidth}
\begin{tabular}{lcc c c c ccc}
\toprule

\multirow{2}{*}{Models}   & \multicolumn{2}{c}{ML-1M}& \multicolumn{2}{c}{BookCrossing}  \\ \cmidrule(lr){2-3}\cmidrule(lr){4-5}
\multirow{2}{*}{}& H@3 $\uparrow$ &  N@3 $\uparrow$ &H@3 $\uparrow$ &  N@3 $\uparrow$  \\ \midrule


LightGCN &0.0283 (-) &0.0203 (-) &0.0358 (-) &0.0272 (-)\\
IFT &0.0268 (-5.30\%) & 0.0193 (-4.93\%) &0.0287 (-19.84\%) & 0.0202 (-25.74\%)\\
\model \textit{w/o} DA &0.0294 (+3.89\%) & 0.0209 (+2.96\%) &0.0408 (+13.97\%) & 0.0319 (+17.28\%)\\
\model \textit{w/o} PA &0.0277 (-2.12\%) & 0.0199 (-1.97\%) &0.0372 (+3.92\%) & 0.0279 (+2.57\%)\\
\model \textit{w/o} AA &0.0298 (+5.30\%) & 0.0218 (+7.39\%) &0.0429 (+19.83\%) & 0.0332 (+22.06\%)\\
\model &\textbf{0.0304 (+7.42\%)} & \textbf{0.0222 (+9.36\%)} &\textbf{0.0434 (+21.23\%)} & \textbf{0.0338 (+24.26\%)}\\

\bottomrule
\end{tabular}
\end{adjustbox}
\vspace{-0.1in}
\end{table}

\subsection{Ablation Study (RQ2)}

We conducted an ablation study to analyze the contributions of different components in our model. Table~\ref{tab:ablation} summarizes the results of the ablation studies across three variants on the ML-1M and Bookcrossing dataset. It is evident that the full model performs considerably better than all its variants, indicating that all the main components contribute significantly to overall performance improvement.
Moreover, compared to the conventional model (\lsc{e.g., LightGCN}), the instruction-tuned LLM (\lsc{e.g., IFT}) does not achieve superior results, underscoring the importance of model aggregation.
We further analyze the specific impact of each component, and our observations are as follows:
\vspace{-0.03in}
\begin{itemize}[leftmargin=*]

\item \textbf{\textit{w/o} Data Augmentation (\textit{w/o} DA)}:
In this variant, we remove the data augmentation module while maintaining other components the same. Experimental results reveal a obvious decline in performance when this module is excluded. This indicates the module's capacity to mitigate data sparsity and long-tail problem, consequently enhancing model performance.

\item \textbf{\textit{w/o} Prompt Augmentation (\textit{w/o} PA)}: In this variant, we remove the prompt augmentation component.
Experimental results demonstrate a significant degradation in model performance when this module is excluded, thereby validating its essential role.
By employing the instruction-tuned LLM with prompt augmentation from prior knowledge by conventional recommendation models, we achieve an enhanced model performance, attributable to the capture of different aspects of information.

\item \textbf{\textit{w/o} Adaptive Aggregation (\textit{w/o} AA)}: In this variant, we substitute adaptive aggregation with uniform aggregation and keep other modules unchanged. Experimental results demonstrate a drop in model performance, underscoring the significance of accounting for the user's long-tail coefficient and employing adaptive aggregation.

\end{itemize}

\subsection{Hyper-parameter Study (RQ3)}
We conducted an analysis of the effects of hyper-parameters $\alpha_1$ and $\alpha_2$. 
These parameters play crucial roles in controlling the weight in adaptive aggregation, as illustrated in Equation \eqref{equ:1}.
Figure \ref{fig:hyp} presents the results on the ML-1M dataset using LightGCN as the backbone model. As $\alpha_1$ increases, we observe an initial surge in the model's performance, followed by a decline. This trend suggests appropriate selection of $\alpha_1$ would enhance the model performance.
With respect to $\alpha_2$, we observe a similar trend but the decline is more pronounced.
This observation is consistent with the principle of adaptive aggregation, which emphasizes the importance of assigning suitable weights to tail users.


\begin{figure}[tbp!]
    \centering
    \begin{minipage}[b]{0.5\linewidth}
        \centering
        \includegraphics[width=\linewidth]{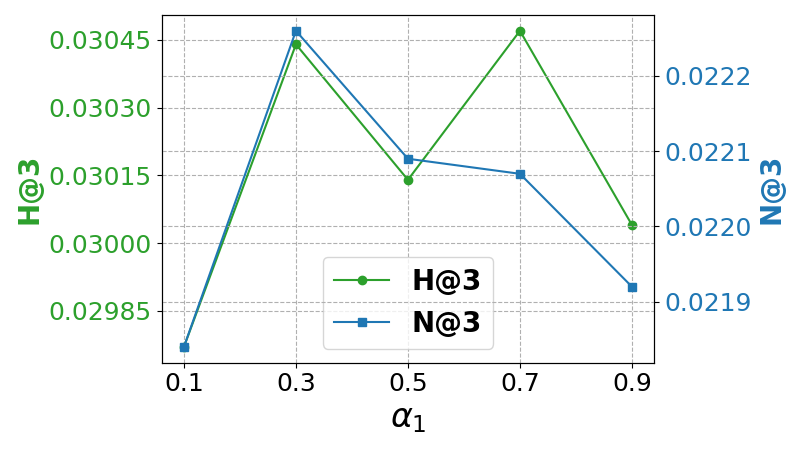}
    \end{minipage}
    \hfill
    \hspace{-10.5pt}
    \begin{minipage}[b]{0.5\linewidth}
        \centering
        \includegraphics[width=\linewidth]{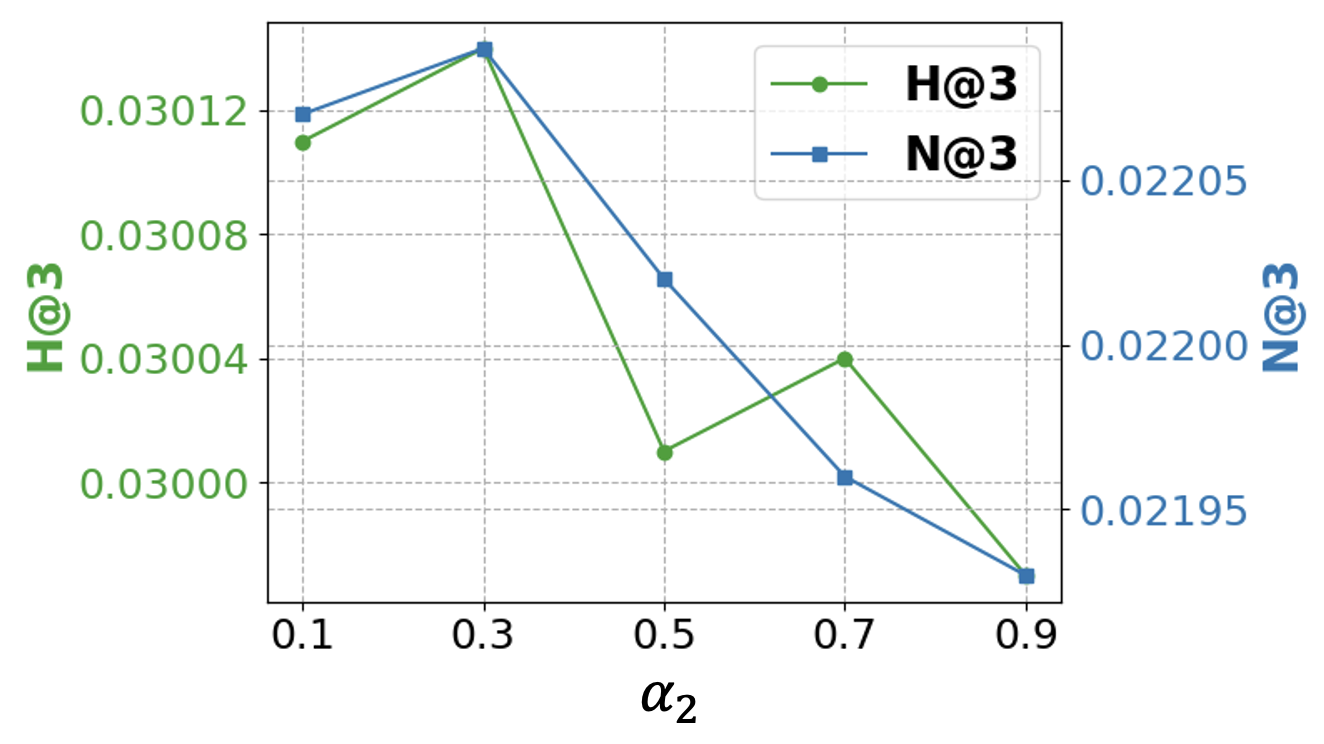}
    \end{minipage}
    \caption{Impact of hyper-parameters $\alpha_1$ and $\alpha_2$ on ML-1M dataset with backbone model LightGCN.}
    \label{fig:hyp}
    \vspace{-0.18in}
\end{figure}

\subsection{Model Analysis (RQ4)}
\begin{figure}[h]
    \centering
    \begin{minipage}[b]{0.5\linewidth}
        \centering
        \includegraphics[width=\linewidth]{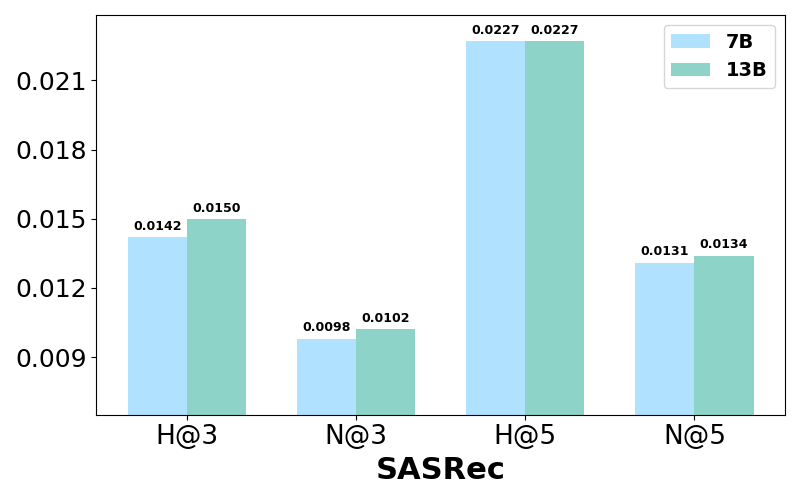}
    \end{minipage}
    \hfill
    \hspace{-10.5pt}
    \begin{minipage}[b]{0.5\linewidth}
        \centering
        \includegraphics[width=\linewidth]{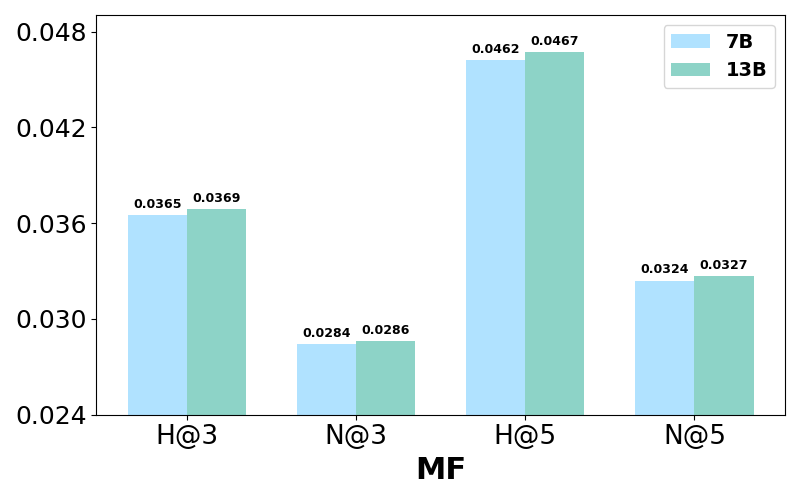}
    \end{minipage}
    
    \caption{Performance comparison w.r.t different LLaMA-2 size for training \model~on the Bookcrossing dataset.
    }
\label{fig:exp_model}
\vspace{-0.05in}
\end{figure} 

\begin{figure}[h]
    \centering
    \begin{minipage}[b]{0.5\linewidth}
        \centering
        \includegraphics[width=\linewidth]{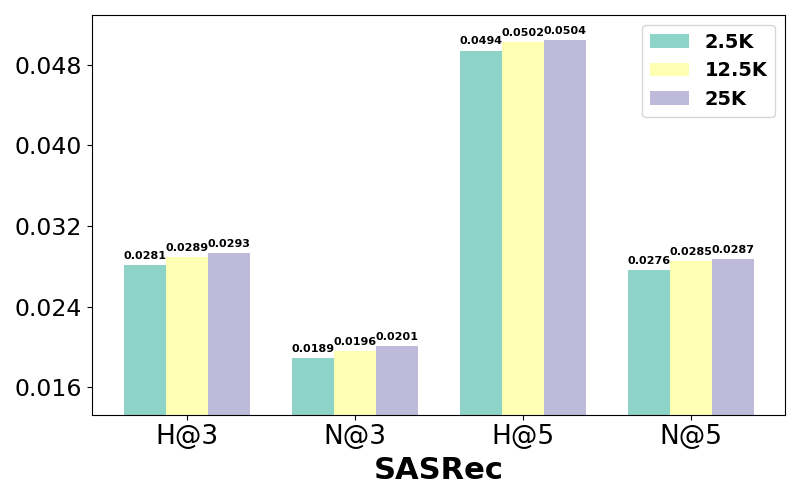}
    \end{minipage}
    \hfill
    \hspace{-10.5pt}
    \begin{minipage}[b]{0.5\linewidth}
        \centering
        \includegraphics[width=\linewidth]{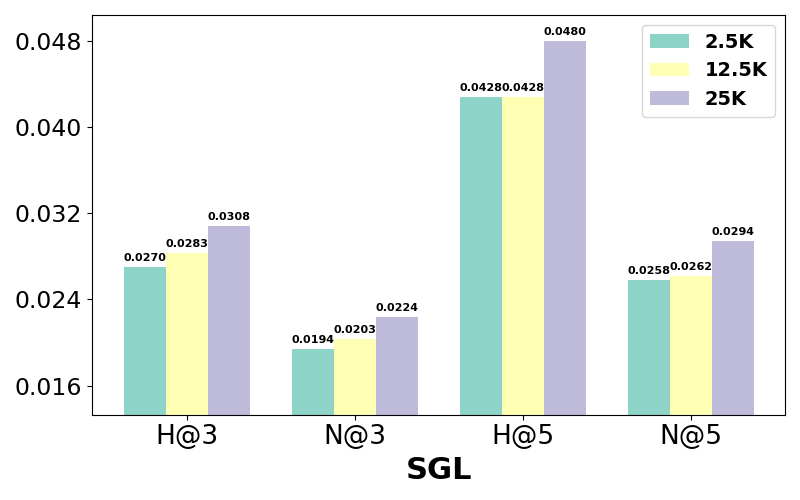}
    \end{minipage}
    
    \caption{Performance comparison w.r.t different numbers of instructions for training \model~on the ML-1M dataset.
    }
\label{fig:exp_data}
\vspace{-0.1in}
\end{figure}




\subsubsection{\textbf{Analysis of Model Scaling}}
We further instruction-tuned the LLaMA-2 model with different model size.\footnote{Due to resource constraints, training the LLaMA-2 (70B) model with identical experimental settings was unfeasible, consistently leading to Out-Of-Memory (OOM) errors.} A comparative analysis was conducted between the 7B and 13B variants of the instruction-tuned models, with performance differences specifically evaluated across various backbone models within the Bookcrossing dataset, as depicted in Figure~\ref{fig:exp_model}. Our findings suggest that the LLaMA-2 (13B) model generally surpasses the 7B version in performance. This can be attributed to the superior language comprehension and reasoning abilities of the larger model, which contribute to improved recommendation results. However, it's worth noting that the improvements are not substantial, indicating that while larger models may provide some performance benefits, the degree of improvement may not always justify the increased computational resources and training time required. It underscores the importance of considering the trade-off between model size, performance gain, and resource efficiency in the design and application of large language models.







\subsubsection{\textbf{Analysis of Data Scaling}}
We evaluated the effect of data size on LLM training by varying the number of instructions in the instruction-tuning dataset. Proportionality with our original configuration, the model with 2.5K instructions underwent 250 training steps, while the 12.5K instructions version was trained over 1250 steps. As depicted in Figure~\ref{fig:exp_data}, a clear trend emerges: model performance improves with an increase in the number of instructions, particularly for direct recommendation models. This highlights the importance of utilizing larger and more diverse datasets for instruction tuning LLMs to optimize performance.







\subsubsection{\textbf{Comparison with the GPT Model}}
We compare our \model with the GPT model, specifically, the GPT-3.5-turbo\footnote{\url{https://platform.openai.com/docs/models/gpt-3-5}} model. 
\lsc{We choose the performance metric hit ratio for comparison.}
A random selection of 100 listwise ranking task samples from the Bookcrossing dataset with backbone model CLSRec served as the benchmark for GPT model.
The experimental result is shown in Figure \ref{fig:comparegpt}.
The experimental result shows \model win over the GPT model in 45\% of the tasks, tied in 25\%, and fell short in 30\% of the times.
This impressive result emphasizes the crucial role of instruction tuning in aligning general-purpose LLMs specifically for recommendation tasks.


\begin{figure}[h]
    \centering
    \includegraphics[width=0.5\linewidth]{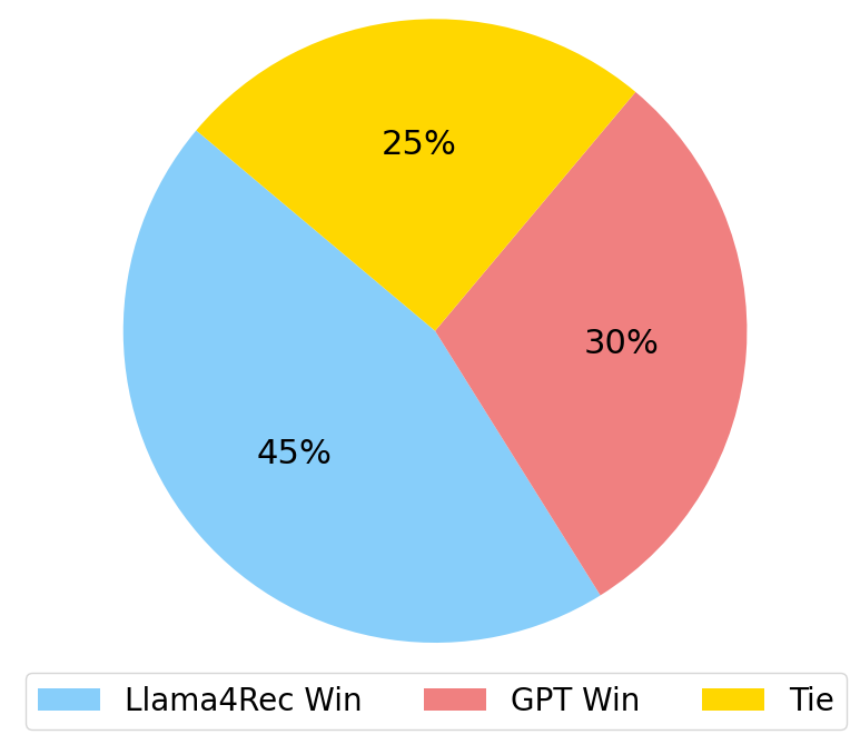}
    \caption{Comparison between our instruction-tuned \model with the GPT-3.5-turbo model. \lsc{Tie refers to cases where the proposed framework and GPT achieve identical performance metrics, specifically the hit ratio.}}
    \label{fig:comparegpt}
\end{figure}

\section{Further Discussion}
\textbf{Model Efficiency.}
This training and inference duration significantly exceeds that of conventional recommendation models, highlighting the limitations of current LLM-based recommender systems. The substantial demand for computational resources also represents a significant challenge. Consequently, employing instruction LLMs for large-scale industrial recommender systems, such as those with millions of users, is presently impractical. However, future advancements in accelerated and parallel computing algorithms for language model inference could potentially reduce inference times and computation resources. This improvement might make the integration of LLMs into large-scale recommender systems feasible, especially by leveraging many GPUs for parallel computation.

\textbf{Linear Combination.}
In this work, we chose a heuristic-based linear interpolation method due to its simplicity and effectiveness, as evidenced by our experiments. Training a neural network to determine optimal aggregation weights would require substantial labeled data to accurately capture the relationship between input factors and performance, which is resource-intensive and impractical for this study. In future work, we plan to explore advanced techniques like genetic programming or reinforcement learning for better modeling of complex factor interactions and enhance nonlinear aggregation effects through dynamic weight optimization.

{\color{black}
\textbf{Hallucination.}
To ensure the accuracy of attributes generated by the LLM, we focused on well-known movies and books, utilizing their reliable metadata. For the evaluation process, we use a two-step approach. First, attributes are generated using the LLaMA model. Then, we employ another advanced LLM, DeepSeek-V3~\cite{liu2024deepseek}, as an evaluator to assess the correctness of the generated attributes. This approach, referred to as the "LLM-as-a-Judge" paradigm~\cite{zheng2023judging}, allows us to systematically evaluate the alignment between generated attributes. 
In cases of disagreement between the evaluator and the generated attributes, we conduct manual evaluations to resolve conflicts, ensuring a higher degree of reliability and accuracy.

To maintain consistency in our analysis, we randomly sampled 50 items from each dataset and utilized greedy decoding during generation to reduce randomness. 
For ML-100K and ML-1M, we evaluated the generated "Movie Director" attribute, while for BookCrossing, we evaluated the generated "Book Genres" attribute.
The quantitative results of this evaluation process are presented in \textbf{Figure~\ref{fig:hall}}.

\begin{figure}[htbp]
    \centering
    \includegraphics[width=0.9\linewidth]{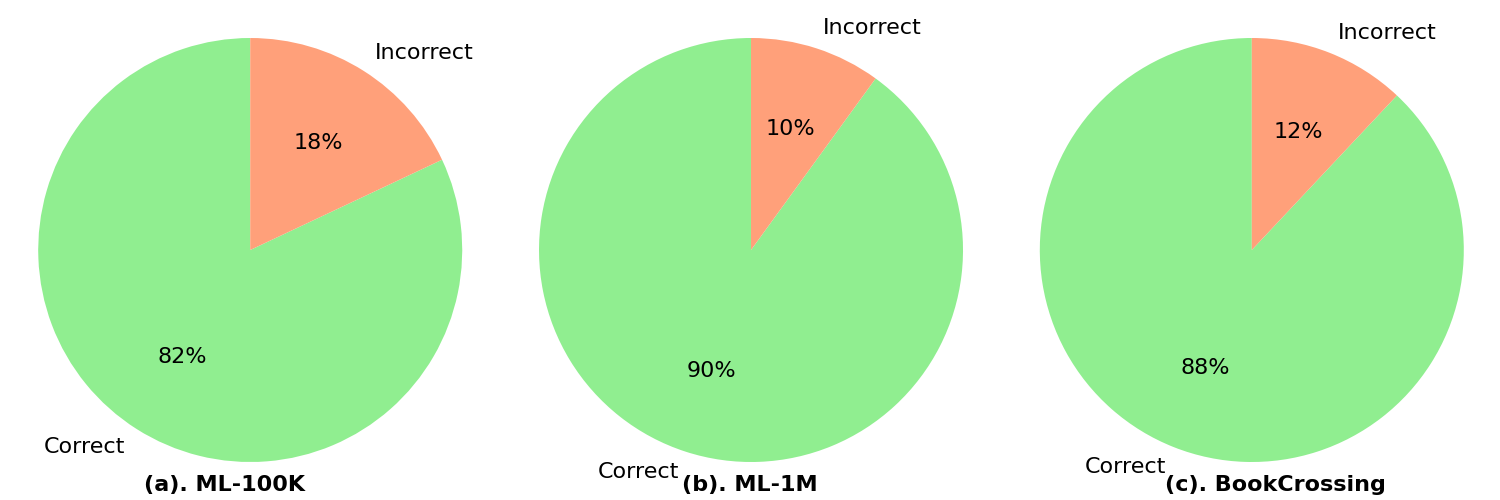}
    \caption{Performance of LLM-generated attributes compared to ground-truth data.}
    \label{fig:hall}
\end{figure}

From \textbf{Figure~\ref{fig:hall}}, we observe that the generated attributes demonstrate a high level of agreement with the ground-truth data, exceeding 80\% across all datasets and reaching of 90\% in ML-1M. 
However, two common types of errors are observed. 
First, in some cases, the Llama model honestly admits it does not know the answer. This behavior could serve as a useful indicator during model training, suggesting that the corresponding movie might not be widely known. 
Second, in other cases, the LLM hallucinates and provides entirely incorrect answers. While these errors are undesirable, they may serve as informative noise, potentially contributing to improved model robustness.

We attribute the high agreement observed to the extensive pretraining of the LLMs on vast amounts of world knowledge, which enables them to generate attributes with a high degree of correctness. However, smaller models are more prone to errors or hallucinations, while larger models generally exhibit better performance due to their increased parameter count, broader training data coverage, and reduced hallucination tendencies.

The world knowledge embedded in LLMs provides a strong foundation for attribute generation. However, to further improve accuracy and mitigate the risk of hallucination, we identify several promising future directions. These include integrating Retrieval-Augmented Generation (RAG) \cite{lewis2020retrieval} to ground the model in external, verifiable knowledge sources, leveraging structured Knowledge Graphs \cite{hogan2021knowledge}, or employing next-generation, more powerful foundation models to mitigate these issues.
}

\section{Conclusion and future work}

In this study, we 
present \model, a general and model-agnostic framework tailored to facilitate mutual augmentation between conventional recommendation models and LLMs through data augmentation and prompt augmentation. Data augmentation for conventional recommendation models could alleviate issues of data sparsity and the long-tail problem, thus improving conventional recommendation model performance. Prompt augmentation, on the other hand, allows the LLM to externalize additional collaborative or sequential information and further enhance the model capability.
Furthermore, adaptive aggregation is employed to merge the predictions from both kinds of augmented models, resulting in more optimized recommendation performance.
Comprehensive experimental results across three diverse recommendation tasks on three real-world datasets demonstrate the effectiveness of \model. While our current approach focuses on mutual augmentation within a single step, our future work will explore expanding mutual augmentation in an iterative manner, potentially unlocking further improvements in model performance.

\section{Acknowledgement}

This work was supported in part by the Research Grants Council of the Hong Kong SAR under Grant GRF 11217823, 11216225, and Collaborative Research Fund C1042-23GF, the National Natural Science Foundation of China under Grant 62371411, InnoHK initiative, the Government of the HKSAR, Laboratory for AI-Powered Financial Technologies.



\bibliographystyle{IEEEtran}
\bibliography{mybib}

@article{bao2023tallrec,
  title={Tallrec: An effective and efficient tuning framework to align large language model with recommendation},
  author={Bao, Keqin and Zhang, Jizhi and Zhang, Yang and Wang, Wenjie and Feng, Fuli and He, Xiangnan},
  journal={arXiv preprint arXiv:2305.00447},
  year={2023}
}

@article{zhang2023recommendation,
  title={Recommendation as instruction following: A large language model empowered recommendation approach},
  author={Zhang, Junjie and Xie, Ruobing and Hou, Yupeng and Zhao, Wayne Xin and Lin, Leyu and Wen, Ji-Rong},
  journal={arXiv preprint arXiv:2305.07001},
  year={2023}
}

@article{liu2023chatgpt,
  title={Is chatgpt a good recommender? a preliminary study},
  author={Liu, Junling and Liu, Chao and Lv, Renjie and Zhou, Kang and Zhang, Yan},
  journal={arXiv preprint arXiv:2304.10149},
  year={2023}
}

@article{zhang2023collm,
  title={Collm: Integrating collaborative embeddings into large language models for recommendation},
  author={Zhang, Yang and Feng, Fuli and Zhang, Jizhi and Bao, Keqin and Wang, Qifan and He, Xiangnan},
  journal={arXiv preprint arXiv:2310.19488},
  year={2023}
}

@inproceedings{geng2022recommendation,
  title={Recommendation as language processing (rlp): A unified pretrain, personalized prompt \& predict paradigm (p5)},
  author={Geng, Shijie and Liu, Shuchang and Fu, Zuohui and Ge, Yingqiang and Zhang, Yongfeng},
  booktitle={Proceedings of the 16th ACM Conference on Recommender Systems},
  pages={299--315},
  year={2022}
}

@article{zheng2023adapting,
  title={Adapting large language models by integrating collaborative semantics for recommendation},
  author={Zheng, Bowen and Hou, Yupeng and Lu, Hongyu and Chen, Yu and Zhao, Wayne Xin and Wen, Ji-Rong},
  journal={arXiv preprint arXiv:2311.09049},
  year={2023}
}

@article{hou2023large,
  title={Large language models are zero-shot rankers for recommender systems},
  author={Hou, Yupeng and Zhang, Junjie and Lin, Zihan and Lu, Hongyu and Xie, Ruobing and McAuley, Julian and Zhao, Wayne Xin},
  journal={arXiv preprint arXiv:2305.08845},
  year={2023}
}

@inproceedings{wu2021self,
  title={Self-supervised graph learning for recommendation},
  author={Wu, Jiancan and Wang, Xiang and Feng, Fuli and He, Xiangnan and Chen, Liang and Lian, Jianxun and Xie, Xing},
  booktitle={Proceedings of the 44th international ACM SIGIR conference on research and development in information retrieval},
  pages={726--735},
  year={2021}
}

@article{fan2023recommender,
  title={Recommender systems in the era of large language models (llms)},
  author={Fan, Wenqi and Zhao, Zihuai and Li, Jiatong and Liu, Yunqing and Mei, Xiaowei and Wang, Yiqi and Tang, Jiliang and Li, Qing},
  journal={arXiv preprint arXiv:2307.02046},
  year={2023}
}

@article{xi2023towards,
  title={Towards Open-World Recommendation with Knowledge Augmentation from Large Language Models},
  author={Xi, Yunjia and Liu, Weiwen and Lin, Jianghao and Zhu, Jieming and Chen, Bo and Tang, Ruiming and Zhang, Weinan and Zhang, Rui and Yu, Yong},
  journal={arXiv preprint arXiv:2306.10933},
  year={2023}
}

@article{anil2023palm,
  title={Palm 2 technical report},
  author={Anil, Rohan and Dai, Andrew M and Firat, Orhan and Johnson, Melvin and Lepikhin, Dmitry and Passos, Alexandre and Shakeri, Siamak and Taropa, Emanuel and Bailey, Paige and Chen, Zhifeng and others},
  journal={arXiv preprint arXiv:2305.10403},
  year={2023}
}

@article{wei2023llmrec,
  title={Llmrec: Large language models with graph augmentation for recommendation},
  author={Wei, Wei and Ren, Xubin and Tang, Jiabin and Wang, Qinyong and Su, Lixin and Cheng, Suqi and Wang, Junfeng and Yin, Dawei and Huang, Chao},
  journal={arXiv preprint arXiv:2311.00423},
  year={2023}
}

@misc{openai2023gpt4,
      title={GPT-4 Technical Report}, 
      author={OpenAI},
      year={2023},
      eprint={2303.08774},
      archivePrefix={arXiv},
      primaryClass={cs.CL}
}

@article{touvron2023llama,
  title={Llama: Open and efficient foundation language models},
  author={Touvron, Hugo and Lavril, Thibaut and Izacard, Gautier and Martinet, Xavier and Lachaux, Marie-Anne and Lacroix, Timoth{\'e}e and Rozi{\`e}re, Baptiste and Goyal, Naman and Hambro, Eric and Azhar, Faisal and others},
  journal={arXiv preprint arXiv:2302.13971},
  year={2023}
}

@article{touvron2023llama2,
  title={Llama 2: Open foundation and fine-tuned chat models},
  author={Touvron, Hugo and Martin, Louis and Stone, Kevin and Albert, Peter and Almahairi, Amjad and Babaei, Yasmine and Bashlykov, Nikolay and Batra, Soumya and Bhargava, Prajjwal and Bhosale, Shruti and others},
  journal={arXiv preprint arXiv:2307.09288},
  year={2023}
}

@article{ouyang2022training,
  title={Training language models to follow instructions with human feedback},
  author={Ouyang, Long and Wu, Jeffrey and Jiang, Xu and Almeida, Diogo and Wainwright, Carroll and Mishkin, Pamela and Zhang, Chong and Agarwal, Sandhini and Slama, Katarina and Ray, Alex and others},
  journal={Advances in Neural Information Processing Systems},
  volume={35},
  pages={27730--27744},
  year={2022}
}

@article{koren2009matrix,
  title={Matrix factorization techniques for recommender systems},
  author={Koren, Yehuda and Bell, Robert and Volinsky, Chris},
  journal={Computer},
  volume={42},
  number={8},
  pages={30--37},
  year={2009},
  publisher={IEEE}
}

@inproceedings{he2017neural,
  title={Neural collaborative filtering},
  author={He, Xiangnan and Liao, Lizi and Zhang, Hanwang and Nie, Liqiang and Hu, Xia and Chua, Tat-Seng},
  booktitle={Proceedings of the 26th international conference on world wide web},
  pages={173--182},
  year={2017}
}

@inproceedings{wang2019neural,
  title={Neural graph collaborative filtering},
  author={Wang, Xiang and He, Xiangnan and Wang, Meng and Feng, Fuli and Chua, Tat-Seng},
  booktitle={Proceedings of the 42nd international ACM SIGIR conference on Research and development in Information Retrieval},
  pages={165--174},
  year={2019}
}

@inproceedings{kang2018self,
  title={Self-attentive sequential recommendation},
  author={Kang, Wang-Cheng and McAuley, Julian},
  booktitle={2018 IEEE international conference on data mining (ICDM)},
  pages={197--206},
  year={2018},
  organization={IEEE}
}

@inproceedings{sun2019bert4rec,
  title={BERT4Rec: Sequential recommendation with bidirectional encoder representations from transformer},
  author={Sun, Fei and Liu, Jun and Wu, Jian and Pei, Changhua and Lin, Xiao and Ou, Wenwu and Jiang, Peng},
  booktitle={Proceedings of the 28th ACM international conference on information and knowledge management},
  pages={1441--1450},
  year={2019}
}

@article{dong2022survey,
  title={A survey for in-context learning},
  author={Dong, Qingxiu and Li, Lei and Dai, Damai and Zheng, Ce and Wu, Zhiyong and Chang, Baobao and Sun, Xu and Xu, Jingjing and Sui, Zhifang},
  journal={arXiv preprint arXiv:2301.00234},
  year={2022}
}

@article{wang2023zero,
  title={Zero-Shot Next-Item Recommendation using Large Pretrained Language Models},
  author={Wang, Lei and Lim, Ee-Peng},
  journal={arXiv preprint arXiv:2304.03153},
  year={2023}
}

@article{rendle2012bpr,
  title={BPR: Bayesian personalized ranking from implicit feedback},
  author={Rendle, Steffen and Freudenthaler, Christoph and Gantner, Zeno and Schmidt-Thieme, Lars},
  journal={arXiv preprint arXiv:1205.2618},
  year={2012}
}

@article{zhang2019deep,
  title={Deep learning based recommender system: A survey and new perspectives},
  author={Zhang, Shuai and Yao, Lina and Sun, Aixin and Tay, Yi},
  journal={ACM computing surveys (CSUR)},
  volume={52},
  number={1},
  pages={1--38},
  year={2019},
  publisher={ACM New York, NY, USA}
}

@inproceedings{park2008long,
  title={The long tail of recommender systems and how to leverage it},
  author={Park, Yoon-Joo and Tuzhilin, Alexander},
  booktitle={Proceedings of the 2008 ACM conference on Recommender systems},
  pages={11--18},
  year={2008}
}

@article{adomavicius2005toward,
  title={Toward the next generation of recommender systems: A survey of the state-of-the-art and possible extensions},
  author={Adomavicius, Gediminas and Tuzhilin, Alexander},
  journal={IEEE transactions on knowledge and data engineering},
  volume={17},
  number={6},
  pages={734--749},
  year={2005},
  publisher={IEEE}
}

@article{hidasi2015session,
  title={Session-based recommendations with recurrent neural networks},
  author={Hidasi, Bal{\'a}zs and Karatzoglou, Alexandros and Baltrunas, Linas and Tikk, Domonkos},
  journal={arXiv preprint arXiv:1511.06939},
  year={2015}
}

@inproceedings{ziegler2005improving,
  title={Improving recommendation lists through topic diversification},
  author={Ziegler, Cai-Nicolas and McNee, Sean M and Konstan, Joseph A and Lausen, Georg},
  booktitle={Proceedings of the 14th international conference on World Wide Web},
  pages={22--32},
  year={2005}
}

@article{harper2015movielens,
  title={The movielens datasets: History and context},
  author={Harper, F Maxwell and Konstan, Joseph A},
  journal={Acm transactions on interactive intelligent systems (tiis)},
  volume={5},
  number={4},
  pages={1--19},
  year={2015},
  publisher={Acm New York, NY, USA}
}

@article{guo2017deepfm,
  title={DeepFM: a factorization-machine based neural network for CTR prediction},
  author={Guo, Huifeng and Tang, Ruiming and Ye, Yunming and Li, Zhenguo and He, Xiuqiang},
  journal={arXiv preprint arXiv:1703.04247},
  year={2017}
}

@incollection{wang2017deep,
  title={Deep \& cross network for ad click predictions},
  author={Wang, Ruoxi and Fu, Bin and Fu, Gang and Wang, Mingliang},
  booktitle={Proceedings of the ADKDD'17},
  pages={1--7},
  year={2017}
}

@inproceedings{lian2018xdeepfm,
  title={xdeepfm: Combining explicit and implicit feature interactions for recommender systems},
  author={Lian, Jianxun and Zhou, Xiaohuan and Zhang, Fuzheng and Chen, Zhongxia and Xie, Xing and Sun, Guangzhong},
  booktitle={Proceedings of the 24th ACM SIGKDD international conference on knowledge discovery \& data mining},
  pages={1754--1763},
  year={2018}
}

@inproceedings{song2019autoint,
  title={Autoint: Automatic feature interaction learning via self-attentive neural networks},
  author={Song, Weiping and Shi, Chence and Xiao, Zhiping and Duan, Zhijian and Xu, Yewen and Zhang, Ming and Tang, Jian},
  booktitle={Proceedings of the 28th ACM international conference on information and knowledge management},
  pages={1161--1170},
  year={2019}
}

@article{xiao2017attentional,
  title={Attentional factorization machines: Learning the weight of feature interactions via attention networks},
  author={Xiao, Jun and Ye, Hao and He, Xiangnan and Zhang, Hanwang and Wu, Fei and Chua, Tat-Seng},
  journal={arXiv preprint arXiv:1708.04617},
  year={2017}
}

@inproceedings{he2020lightgcn,
  title={Lightgcn: Simplifying and powering graph convolution network for recommendation},
  author={He, Xiangnan and Deng, Kuan and Wang, Xiang and Li, Yan and Zhang, Yongdong and Wang, Meng},
  booktitle={Proceedings of the 43rd International ACM SIGIR conference on research and development in Information Retrieval},
  pages={639--648},
  year={2020}
}

@inproceedings{xie2022contrastive,
  title={Contrastive learning for sequential recommendation},
  author={Xie, Xu and Sun, Fei and Liu, Zhaoyang and Wu, Shiwen and Gao, Jinyang and Zhang, Jiandong and Ding, Bolin and Cui, Bin},
  booktitle={2022 IEEE 38th international conference on data engineering (ICDE)},
  pages={1259--1273},
  year={2022},
  organization={IEEE}
}

@inproceedings{rajbhandari2020zero,
  title={Zero: Memory optimizations toward training trillion parameter models},
  author={Rajbhandari, Samyam and Rasley, Jeff and Ruwase, Olatunji and He, Yuxiong},
  booktitle={SC20: International Conference for High Performance Computing, Networking, Storage and Analysis},
  pages={1--16},
  year={2020},
  organization={IEEE}
}

@article{dao2022flashattention,
  title={Flashattention: Fast and memory-efficient exact attention with io-awareness},
  author={Dao, Tri and Fu, Dan and Ermon, Stefano and Rudra, Atri and R{\'e}, Christopher},
  journal={Advances in Neural Information Processing Systems},
  volume={35},
  pages={16344--16359},
  year={2022}
}

@misc{alpaca,
  author = {Rohan Taori and Ishaan Gulrajani and Tianyi Zhang and Yann Dubois and Xuechen Li and Carlos Guestrin and Percy Liang and Tatsunori B. Hashimoto },
  title = {Stanford Alpaca: An Instruction-following LLaMA model},
  year = {2023},
  publisher = {GitHub},
  journal = {GitHub repository},
  howpublished = {\url{https://github.com/tatsu-lab/stanford_alpaca}},
}

@inproceedings{fan2019graph,
  title={Graph neural networks for social recommendation},
  author={Fan, Wenqi and Ma, Yao and Li, Qing and He, Yuan and Zhao, Eric and Tang, Jiliang and Yin, Dawei},
  booktitle={The world wide web conference},
  pages={417--426},
  year={2019}
}

@inproceedings{luo2022hysage,
  title={HySAGE: A hybrid static and adaptive graph embedding network for context-drifting recommendations},
  author={Luo, Sichun and Zhang, Xinyi and Xiao, Yuanzhang and Song, Linqi},
  booktitle={Proceedings of the 31st ACM International Conference on Information \& Knowledge Management},
  pages={1389--1398},
  year={2022}
}

@article{yu2023self,
  title={Self-supervised learning for recommender systems: A survey},
  author={Yu, Junliang and Yin, Hongzhi and Xia, Xin and Chen, Tong and Li, Jundong and Huang, Zi},
  journal={IEEE Transactions on Knowledge and Data Engineering},
  year={2023},
  publisher={IEEE}
}

@article{khan2021deep,
  title={Deep learning techniques for rating prediction: a survey of the state-of-the-art},
  author={Khan, Zahid Younas and Niu, Zhendong and Sandiwarno, Sulis and Prince, Rukundo},
  journal={Artificial Intelligence Review},
  volume={54},
  pages={95--135},
  year={2021},
  publisher={Springer}
}

@inproceedings{steck2013evaluation,
  title={Evaluation of recommendations: rating-prediction and ranking},
  author={Steck, Harald},
  booktitle={Proceedings of the 7th ACM conference on Recommender systems},
  pages={213--220},
  year={2013}
}

@article{le2021efficient,
  title={Efficient retrieval of matrix factorization-based top-k recommendations: A survey of recent approaches},
  author={Le, Dung D and Lauw, Hady},
  journal={Journal of Artificial Intelligence Research},
  volume={70},
  pages={1441--1479},
  year={2021}
}

@inproceedings{yang2012top,
  title={On top-k recommendation using social networks},
  author={Yang, Xiwang and Steck, Harald and Guo, Yang and Liu, Yong},
  booktitle={Proceedings of the sixth ACM conference on Recommender systems},
  pages={67--74},
  year={2012}
}

@inproceedings{huang2021mixgcf,
  title={Mixgcf: An improved training method for graph neural network-based recommender systems},
  author={Huang, Tinglin and Dong, Yuxiao and Ding, Ming and Yang, Zhen and Feng, Wenzheng and Wang, Xinyu and Tang, Jie},
  booktitle={Proceedings of the 27th ACM SIGKDD Conference on Knowledge Discovery \& Data Mining},
  pages={665--674},
  year={2021}
}

@article{longpre2023flan,
  title={The flan collection: Designing data and methods for effective instruction tuning},
  author={Longpre, Shayne and Hou, Le and Vu, Tu and Webson, Albert and Chung, Hyung Won and Tay, Yi and Zhou, Denny and Le, Quoc V and Zoph, Barret and Wei, Jason and others},
  journal={arXiv preprint arXiv:2301.13688},
  year={2023}
}

@article{zhang2023instruction,
  title={Instruction tuning for large language models: A survey},
  author={Zhang, Shengyu and Dong, Linfeng and Li, Xiaoya and Zhang, Sen and Sun, Xiaofei and Wang, Shuhe and Li, Jiwei and Hu, Runyi and Zhang, Tianwei and Wu, Fei and others},
  journal={arXiv preprint arXiv:2308.10792},
  year={2023}
}

@article{chen2023sim2rec,
  title={Sim2Rec: A Simulator-based Decision-making Approach to Optimize Real-World Long-term User Engagement in Sequential Recommender Systems},
  author={Chen, Xiong-Hui and He, Bowei and Yu, Yang and Li, Qingyang and Qin, Zhiwei and Shang, Wenjie and Ye, Jieping and Ma, Chen},
  journal={arXiv preprint arXiv:2305.04832},
  year={2023}
}

@article{luo2024recranker,
  title={Recranker: Instruction tuning large language model as ranker for top-k recommendation},
  author={Luo, Sichun and He, Bowei and Zhao, Haohan and Shao, Wei and Qi, Yanlin and Huang, Yinya and Zhou, Aojun and Yao, Yuxuan and Li, Zongpeng and Xiao, Yuanzhang and others},
  journal={ACM Transactions on Information Systems},
  year={2024},
  publisher={ACM New York, NY}
}

@article{doppala2022reliable,
  title={A reliable machine intelligence model for accurate identification of cardiovascular diseases using ensemble techniques},
  author={Doppala, Bhanu Prakash and Bhattacharyya, Debnath and Janarthanan, Midhunchakkaravarthy and Baik, Namkyun},
  journal={Journal of Healthcare Engineering},
  volume={2022},
  number={1},
  pages={2585235},
  year={2022},
  publisher={Wiley Online Library}
}

@article{zounemat2021ensemble,
  title={Ensemble machine learning paradigms in hydrology: A review},
  author={Zounemat-Kermani, Mohammad and Batelaan, Okke and Fadaee, Marzieh and Hinkelmann, Reinhard},
  journal={Journal of Hydrology},
  volume={598},
  pages={126266},
  year={2021},
  publisher={Elsevier}
}

@inproceedings{luo2022personalized,
  title={Personalized federated recommendation via joint representation learning, user clustering, and model adaptation},
  author={Luo, Sichun and Xiao, Yuanzhang and Song, Linqi},
  booktitle={Proceedings of the 31st ACM international conference on information \& knowledge management},
  pages={4289--4293},
  year={2022}
}

@article{luo2024perfedrec++,
  title={Perfedrec++: Enhancing personalized federated recommendation with self-supervised pre-training},
  author={Luo, Sichun and Xiao, Yuanzhang and Zhang, Xinyi and Liu, Yang and Ding, Wenbo and Song, Linqi},
  journal={ACM Transactions on Intelligent Systems and Technology},
  volume={15},
  number={5},
  pages={1--24},
  year={2024},
  publisher={ACM New York, NY, USA}
}

@inproceedings{luo2023improving,
  title={Improving long-tail item recommendation with graph augmentation},
  author={Luo, Sichun and Ma, Chen and Xiao, Yuanzhang and Song, Linqi},
  booktitle={Proceedings of the 32nd ACM international conference on information and knowledge management},
  pages={1707--1716},
  year={2023}
}

@article{luo2025reasoning,
  title={Reasoning Meets Personalization: Unleashing the Potential of Large Reasoning Model for Personalized Generation},
  author={Luo, Sichun and Deng, Guanzhi and Xu, Jian and Zhang, Xiaojie and Hou, Hanxu and Song, Linqi},
  journal={arXiv preprint arXiv:2505.17571},
  year={2025}
}

@article{luo2025rallrec+,
  title={RALLRec+: Retrieval Augmented Large Language Model Recommendation with Reasoning},
  author={Luo, Sichun and Xu, Jian and Zhang, Xiaojie and Wang, Linrong and Liu, Sicong and Hou, Hanxu and Song, Linqi},
  journal={arXiv preprint arXiv:2503.20430},
  year={2025}
}

@article{luo2024privacy,
  title={Privacy in LLM-based Recommendation: Recent Advances and Future Directions},
  author={Luo, Sichun and Shao, Wei and Yao, Yuxuan and Xu, Jian and Liu, Mingyang and Li, Qintong and He, Bowei and Wang, Maolin and Deng, Guanzhi and Hou, Hanxu and others},
  journal={arXiv preprint arXiv:2406.01363},
  year={2024}
}

@inproceedings{kwon2023efficient,
  title={Efficient memory management for large language model serving with pagedattention},
  author={Kwon, Woosuk and Li, Zhuohan and Zhuang, Siyuan and Sheng, Ying and Zheng, Lianmin and Yu, Cody Hao and Gonzalez, Joseph and Zhang, Hao and Stoica, Ion},
  booktitle={Proceedings of the 29th Symposium on Operating Systems Principles},
  pages={611--626},
  year={2023}
}

@inproceedings{luo2024large,
  title={Large language models augmented rating prediction in recommender system},
  author={Luo, Sichun and Wang, Jiansheng and Zhou, Aojun and Ma, Li and Song, Linqi},
  booktitle={ICASSP 2024-2024 IEEE International Conference on Acoustics, Speech and Signal Processing (ICASSP)},
  pages={7960--7964},
  year={2024},
  organization={IEEE}
}

@article{lin2025can,
  title={How can recommender systems benefit from large language models: A survey},
  author={Lin, Jianghao and Dai, Xinyi and Xi, Yunjia and Liu, Weiwen and Chen, Bo and Zhang, Hao and Liu, Yong and Wu, Chuhan and Li, Xiangyang and Zhu, Chenxu and others},
  journal={ACM Transactions on Information Systems},
  volume={43},
  number={2},
  pages={1--47},
  year={2025},
  publisher={ACM New York, NY}
}

@article{zhao2024recommender,
  title={Recommender systems in the era of large language models (llms)},
  author={Zhao, Zihuai and Fan, Wenqi and Li, Jiatong and Liu, Yunqing and Mei, Xiaowei and Wang, Yiqi and Wen, Zhen and Wang, Fei and Zhao, Xiangyu and Tang, Jiliang and others},
  journal={IEEE Transactions on Knowledge and Data Engineering},
  year={2024},
  publisher={IEEE}
}

@inproceedings{dai2023uncovering,
  title={Uncovering chatgpt’s capabilities in recommender systems},
  author={Dai, Sunhao and Shao, Ninglu and Zhao, Haiyuan and Yu, Weijie and Si, Zihua and Xu, Chen and Sun, Zhongxiang and Zhang, Xiao and Xu, Jun},
  booktitle={Proceedings of the 17th ACM Conference on Recommender Systems},
  pages={1126--1132},
  year={2023}
}

@inproceedings{wei2024llmrec,
  title={Llmrec: Large language models with graph augmentation for recommendation},
  author={Wei, Wei and Ren, Xubin and Tang, Jiabin and Wang, Qinyong and Su, Lixin and Cheng, Suqi and Wang, Junfeng and Yin, Dawei and Huang, Chao},
  booktitle={Proceedings of the 17th ACM International Conference on Web Search and Data Mining},
  pages={806--815},
  year={2024}
}

@inproceedings{xi2024towards,
  title={Towards open-world recommendation with knowledge augmentation from large language models},
  author={Xi, Yunjia and Liu, Weiwen and Lin, Jianghao and Cai, Xiaoling and Zhu, Hong and Zhu, Jieming and Chen, Bo and Tang, Ruiming and Zhang, Weinan and Yu, Yong},
  booktitle={Proceedings of the 18th ACM Conference on Recommender Systems},
  pages={12--22},
  year={2024}
}

@inproceedings{wang2024large,
  title={Large language models as data augmenters for cold-start item recommendation},
  author={Wang, Jianling and Lu, Haokai and Caverlee, James and Chi, Ed H and Chen, Minmin},
  booktitle={Companion Proceedings of the ACM Web Conference 2024},
  pages={726--729},
  year={2024}
}

@inproceedings{zheng2024adapting,
  title={Adapting large language models by integrating collaborative semantics for recommendation},
  author={Zheng, Bowen and Hou, Yupeng and Lu, Hongyu and Chen, Yu and Zhao, Wayne Xin and Chen, Ming and Wen, Ji-Rong},
  booktitle={2024 IEEE 40th International Conference on Data Engineering (ICDE)},
  pages={1435--1448},
  year={2024},
  organization={IEEE}
}

@inproceedings{ren2024representation,
  title={Representation learning with large language models for recommendation},
  author={Ren, Xubin and Wei, Wei and Xia, Lianghao and Su, Lixin and Cheng, Suqi and Wang, Junfeng and Yin, Dawei and Huang, Chao},
  booktitle={Proceedings of the ACM Web Conference 2024},
  pages={3464--3475},
  year={2024}
}

@article{lewis2020retrieval,
  title={Retrieval-augmented generation for knowledge-intensive nlp tasks},
  author={Lewis, Patrick and Perez, Ethan and Piktus, Aleksandra and Petroni, Fabio and Karpukhin, Vladimir and Goyal, Naman and K{\"u}ttler, Heinrich and Lewis, Mike and Yih, Wen-tau and Rockt{\"a}schel, Tim and others},
  journal={Advances in neural information processing systems},
  volume={33},
  pages={9459--9474},
  year={2020}
}

@article{zheng2023judging,
  title={Judging llm-as-a-judge with mt-bench and chatbot arena},
  author={Zheng, Lianmin and Chiang, Wei-Lin and Sheng, Ying and Zhuang, Siyuan and Wu, Zhanghao and Zhuang, Yonghao and Lin, Zi and Li, Zhuohan and Li, Dacheng and Xing, Eric and others},
  journal={Advances in Neural Information Processing Systems},
  volume={36},
  pages={46595--46623},
  year={2023}
}

@article{liu2024deepseek,
  title={Deepseek-v3 technical report},
  author={Liu, Aixin and Feng, Bei and Xue, Bing and Wang, Bingxuan and Wu, Bochao and Lu, Chengda and Zhao, Chenggang and Deng, Chengqi and Zhang, Chenyu and Ruan, Chong and others},
  journal={arXiv preprint arXiv:2412.19437},
  year={2024}
}

@article{liu2024lost,
  title={Lost in the Middle: How Language Models Use Long Contexts},
  author={Liu, Nelson F and Lin, Kevin and Hewitt, John and Paranjape, Ashwin and Bevilacqua, Michele and Petroni, Fabio and Liang, Percy},
  journal={Transactions of the Association for Computational Linguistics},
  volume={12},
  pages={157--173},
  year={2024},
  publisher={MIT Press}
}

@article{hogan2021knowledge,
  title={Knowledge graphs},
  author={Hogan, Aidan and Blomqvist, Eva and Cochez, Michael and d’Amato, Claudia and Melo, Gerard De and Gutierrez, Claudio and Kirrane, Sabrina and Gayo, Jos{\'e} Emilio Labra and Navigli, Roberto and Neumaier, Sebastian and others},
  journal={ACM Computing Surveys (Csur)},
  volume={54},
  number={4},
  pages={1--37},
  year={2021},
  publisher={ACM New York, NY, USA}
}

\end{document}